\documentclass[pre,twocolumn,showpacs,eqsecnum,amsmath,amssymb]{revtex4}
\usepackage{amsmath,amscd,amsthm,amssymb,mathrsfs,amsfonts}
\usepackage[colorlinks]{hyperref}
\usepackage{graphicx}
\usepackage{bm}

\begin{document}

\title{Relativistic Lagrangian model of a nematic liquid crystal interacting with an electromagnetic field}

\author{\firstname{Yuri N.}~\surname{Obukhov}}
\email{yo@thp.uni-koeln.de}
\affiliation{Institute for Theoretical Physics, University of Cologne, 
50923 K\"oln, Germany}

\author{\firstname{Tom\'as}~\surname{Ramos}}
\email{tramos@udec.cl}
\affiliation{Departamento de F\'isica, Universidad de Concepci\'on, Casilla 160-C, Concepci\'on, Chile}
\affiliation{Institute for Theoretical Physics, University of Innsbruck, A-6020 Innsbruck, Austria}
\affiliation{Institute for Quantum Optics and Quantum Information of the Austrian Academy of Sciences, A-6020 Innsbruck, Austria}

\author{\firstname{Guillermo F.}~\surname{Rubilar}} 
\email{grubilar@udec.cl}
\affiliation{Departamento de F\'isica, Universidad de Concepci\'on, Casilla 160-C, Concepci\'on, Chile}

\date{\today}
\begin{abstract}
We develop a relativistic variational model for a nematic liquid crystal interacting with an electromagnetic field. 
The constitutive relation for a general anisotropic uniaxial diamagnetic and dielectric 
medium is analyzed. 
We discuss light wave propagation in this moving uniaxial medium, for which the corresponding optical metrics are identified explicitly.
A Lagrangian for the coupled system of a nematic liquid crystal and the electromagnetic field is constructed, from which a complete set of equations of motion for the system is derived. The canonical energy-momentum and spin tensors are systematically obtained. 
We compare our results with those within the non-relativistic models. 
As an application of our general formalism, we discuss the so-called 
Abraham-Minkowski controversy on the momentum of light in a medium.
\end{abstract}
\pacs{03.50.De; 47.75.+f; 83.80.Xz; 47.10.ab; 03.50.-z}
\maketitle

\section{Introduction}

Liquid crystals provide an interesting example of a subject where the
fundamental and applied sciences are deeply related. After their first
experimental discovery more than 120 years ago, great number of substances 
(natural and synthesized) with the properties of liquid crystals are known, 
which have many important practical applications for modern technology. 
Good overviews and introduction to this subject can be found, for example,
in \cite{degennes,chandra,dejeu,blinov,demus,scharf,Virga,stewart}. 
In our study, we deal with nematic liquid crystals (with a possible 
generalization to cholesteric crystals) which fall into a particular class of media
with microstructure. In classical continuum mechanics, a material medium 
consists of structureless points. In the early 20th century, 
the Cosserat brothers \cite{coss} proposed a generalization of this simple picture, in which the material body or fluid is formed by particles whose microscopic properties contribute to the macroscopic dynamics of the medium. These more complex continuous mechanical models are known under different names such as the theories of multipolar, micromorphic or oriented media \cite{capriz,Eringen99,TT}. An important 
particular case of continua with microstructure is represented by the
spinning fluids \cite{WR,halb,corben}. 

As a medium with anisotropic electromagnetic (optical) properties, a nematic liquid crystal is another particular case of a medium with microstructure. Just like the spinning fluid, which is characterized by elements with an ``internal'' degree of freedom associated with spin, the liquid crystal is a medium of ``stretched'' particles whose orientational motion is described by an additional hydrodynamic variable. Mathematically, this additional degree of freedom is a unit vector field ${\bm n}={\bm n}({\bm x},t),\ {\bm n}^2=1$, which is called {\it director}.
The director is a microscopic variable that is assigned to every material 
point of the medium. For the cholesteric liquid crystals, in addition, the 
chirality (handedness) property is assigned to the material points. 

We construct here a complete relativistic Lagrangian theory of nematic
liquid crystals interacting with the electromagnetic field. The non-relativistic
variational models were developed previously in \cite{lisin1,lisin2}, see
also \cite{Virga}. This variational approach is convenient for the
study of the full nonlinear dynamics of a liquid crystal, namely the equations
of motion and the conservation laws. Relativistic fluid models are working 
tools in various fields of research such as
high-energy plasma astrophysics and nuclear physics (where non-ideal fluids 
are extremely successfully applied to the description of heavy ion reactions) 
\cite{anile1,anile2}. Also in cosmology, hydrodynamical descriptions of matter 
are standard both for the early and for the later stages of the evolution of 
the Universe. Our derivations make use of the earlier studies in which the
relativistic Lagrangian theories were developed for ideal fluids without 
\cite{taub1,schutz,ray72} and with microstructure \cite{seliger}. Of special interest are the models \cite{spinfluid1,spinfluid2} of relativistic spinning fluids. Since we use the Lagrangian formalism it is nontrivial to take into account dissipative effects. In this work we neglect dissipation in the motion of the relativistic liquid crystal and therefore our model has to be understood as a first step towards a more realistic theory. On the other hand, the description of light propagation in this medium is not affected by the inclusion of viscosity.  Additionally, even in the case in which the medium is a ``rigid'' non-dissipative anisotropic (birefringent) crystal, a  model in terms of a ``liquid'' (i.e. a fluid) is needed to consistently describe its response to the electromagnetic field, since the concept of rigid bodies is incompatible with a relativistic description.

The general framework of the latter works has been used recently for the 
investigation of the problem of the energy and momentum of the electromagnetic
field in moving media \cite{obukhov1,dielectricslab}. However, this study has been restricted to isotropic media only. Here we further develop a relativistic Lagrangian fluid model and apply the extended theory to the
description of a nematic liquid crystal and its interaction with the 
electromagnetic field. This is an example of light interacting with a dynamical anisotropic medium, which can be used to gain a deeper insight in the long standing Abraham-Minkowski controversy \cite{penfieldhaus1,penfieldhaus1967,pfeifer,barnett2010,barnettloudon2,milonniboyd,Griffiths}.

As a starting point we use the nonrelativistic model \cite{lisin1,lisin2}. The important difference is however that we use the
Euler picture and not the Lagrange one as in \cite{lisin1,lisin2}. The 
Euler approach seems to be more physically attractive since it provides the
field-theoretical treatment for both the electromagnetic field and the 
material variables that describe the dynamics of the liquid crystal. 

Our discussion here is confined to the flat Minkowski spacetime. In other
words, we do not consider the gravitational effects encoded in a 
nontrivial Riemannian metric. However, the generalization of the relativistic 
Lagrangian model of liquid crystals to curved spacetimes is fairly
straightforward. A possible physically important application of such a
generalization would be a reanalysis of the structure and defect formation
(walls, strings, textures) in cosmology, relying on the analogy between
cosmological defects and defects in liquid crystals \cite{cosmo1,cosmo2}.

Our notation follows \cite{obukhov1} and the book \cite{birk}. In particular, the indices from the middle of the Latin alphabet $i,j,k,\ldots 
= 0,1,2,3$ label the 4-dimensional spacetime components, the Latin indices from the beginning of the alphabet $a,b,c,\ldots = 1,2,3$ refer to the 3-dimensional 
spatial objects and operations (the 3-vectors are also displayed in boldface). The Minkowski metric is defined as $g_{ij}:={\rm diag}(c^2,-1,-1,-1)$ and the 3-dimensional Levi-Civita symbol in the rest frame is chosen so that ${\stackrel {\circ}\epsilon}{}^{123}:=1$. The determinant is denoted as usual by $g = {\rm det}(\,g_{ij})$; thus in Minkowski spacetime $\sqrt{-g} = c$.  
In the 4-dimensional framework spatial components of tensor must be raised or lowered by $g_{ab}=-\delta_{ab}$, but when we are working only with 3-dimensional tensors, we use the convention of using just the Euclidean metric $\delta_{ab}$ to raise and lower spatial indices.

\section{Constitutive relations of a liquid crystal at rest}\label{constirest}

In Maxwell's theory \cite{birk} the electromagnetic field is described 
by the electromagnetic field strength $F_{ij} = (\bm{E},\bm{B})$ and the electromagnetic
excitation $H^{ij} =(\bm{D},\bm{H})$. Additionally, in order to obtain a predictive theory, 
one needs to specify the constitutive relations $H^{ij} = H^{ij}(F_{kl})$ for the specific 
medium under consideration.

A liquid crystal is a medium with uniaxial anisotropic properties. The 
constitutive relations for a medium with electric and magnetic
properties of this kind can be expressed, when the medium is at rest, by
\begin{align}
D^a ={}&\varepsilon_0\,{\stackrel \circ \varepsilon}{}^{ab}E_b,\label{c1}\\
H_a ={}&\mu_0^{-1}\,{\stackrel \circ \mu}{}^{-1}_{ab}B^b.\label{c2}
\end{align}
Here ${\stackrel \circ\varepsilon}{}^{ab}$ is the relative
permittivity tensor and ${\stackrel \circ\mu}{}^{-1}_{ab}$ is the
inverse of the relative permeability tensor ${\stackrel \circ\mu}{}^{ab}$. 
The fact that all quantities are considered in the frame where the medium 
is at rest is indicated with the symbol ${}^\circ$. 

Alternatively to (\ref{c1}) and (\ref{c2}), it is also useful
to write the constitutive relations in terms of the polarization and magnetization fields:
\begin{align}
P^a={}&\varepsilon_0 ({\stackrel \circ \varepsilon}{}^{ab} - \delta^{ab})E_b,\label{Pa}\\
M_a={}&\mu^{-1}_0(\delta_{ab} - {\stackrel \circ \mu}{}^{-1}_{ab})B^b.\label{Ma}
\end{align}

In the case when the medium has the same optical axis $\bm{n}$ for the electric and magnetic anisotropy, it is convenient to decompose ${\stackrel \circ\varepsilon}{}^{ab}$ and ${\stackrel \circ\mu}{}^{-1}_{ab}$ in terms of its eigenvectors.  By choosing these vectors as $\bm{n}$, $\bm{n}_1$, and $\bm{n}_2$, with two eigenvalues equal and one different, we have
\begin{align}
{\stackrel \circ\varepsilon}{}^{ab}={}&\varepsilon_{\|}n^an^b
+\varepsilon_{\perp}(n^a_1n^b_1+n^a_2n^b_2),\label{epsilondecom}\\
{\stackrel \circ\mu}{}^{-1}_{ab}={}&\mu^{-1}_{\|}n_an_b
+\mu^{-1}_{\perp}(n^1_{a}n^1_{b}+n^2_{a}n^2_{b}).\label{invmudecom}
\end{align}
Here $\varepsilon_{\perp}$ and $\varepsilon_{\|}$ are the relative permittivities, perpendicular and parallel to the optical axis vector $\bm{n}$ (with Cartesian components $n^a$); $\mu_{\perp}$ and $\mu_{\|}$ are the corresponding perpendicular and parallel relative permeability
functions. Since the tensors ${\stackrel \circ\varepsilon}{}^{ab}$ and ${\stackrel \circ\mu}{}^{-1}_{ab}$ are symmetric 
and real for a non-dissipative medium, the eigenvectors satisfy the closure relation
\begin{align}
n^an^b + n^a_1n^b_1 + n^a_2n^b_2 = \delta^{ab}.\label{closure1}
\end{align}
Inserting (\ref{closure1}) in (\ref{epsilondecom}) and (\ref{invmudecom}), 
we obtain a simpler expression for the dielectric and magnetic tensors when the medium is at rest,
\begin{align}
{\stackrel \circ\varepsilon}{}^{ab}={}&\varepsilon_{\perp}\,\delta^{ab}+\Delta\varepsilon\,n^an^b,\label{epij}\\
{\stackrel \circ\mu}{}^{-1}_{ab}={}&\mu^{-1}_{\perp}\,\delta_{ab}+\Delta\mu^{-1}\,n_an_b. \label{invmuij}
\end{align}
The dielectric and magnetic anisotropies are defined, respectively, by
\begin{align}
\Delta\varepsilon :={}&\varepsilon_{\|}-\varepsilon_{\perp},\label{deps}\\
\Delta\mu^{-1}:={}&\mu^{-1}_{\|}-\mu^{-1}_{\perp}.\label{dmu}
\end{align}
The latter quantity should {\it not} be misunderstood as the inverse of the
difference $\mu_{\|}-\mu_{\perp}$, that is, $\Delta\mu^{-1} \neq (\Delta\mu)^{-1}$.
Strictly speaking, one should write (\ref{dmu}) as $\Delta(\mu^{-1})$, but we omit the 
parentheses to simplify the formulas. 

\section{Non-relativistic Lagrangian for a nematic liquid crystal}

The non-relativistic liquid crystal theory is a well established subject;
see, for instance,  \cite{degennes,dejeu,stewart,erik5,erik6,leslie1,leslie2,erik2,erik3,erik4}. Nevertheless, the Lagrangian 
approach for the study of the dynamics of this medium with microstructure was developed only recently in \cite{lisin1,lisin2} 
(although the variational methods were used in \cite{Virga} for the analysis of the equilibrium 
problems for the liquid crystals). 

According to \cite{lisin1,lisin2}, the non-relativistic kinetic energy density of a liquid crystal reads
\begin{align}
{\cal K}:=\rho_{\rm m}\frac{{\bm v}^2}{2}+\rho_{\rm m} {\cal J} \frac{{\bm \omega}^2}{2}.\label{kineticnonrel}
\end{align}
Here $\rho_{\rm m}({\bm x},t)$ is the mass density of the liquid crystal, ${\bm v}({\bm x},t)$ its velocity field, $\cal J$ the geometric moment of inertia of a fluid element (dimensionless), and ${\bm \omega}$ the angular velocity of the director, which is defined as
\begin{align}
{\bm \omega}:={\bm n}\times\dot{{\bm n}},\label{omegaa}
\end{align}
with ${\bm n}({\bm x},t)$ the director field of the liquid crystal and
\begin{align}
\dot{{\bm n}}:=\frac{\partial{\bm n}}{\partial t}+({\bm v}\cdot{\bm \nabla}){\bm n},
\end{align}
the convective derivative of the director.

The potential energy density is represented by the free energy ${\cal F}$, which is usually taken as the thermodynamic potential in the theory of liquid crystals. Following \cite{lisin1,lisin2,degennes,stewart}, we express the free energy as
\begin{align}\label{ff}
{\cal F} = {\cal F}_0 + {\cal F}_{\rm d} + {\cal F}_{\rm e} + {\cal F}_{\rm m},
\end{align} 
where ${\cal F}_0={\cal F}_0(\rho_{\rm m},T)$ is the internal free
energy which describes the hydrodynamic portion of ${\cal F}$ and
depends on the density $\rho_{\rm m}$ and the 
temperature $T$. The pressure in the medium is introduced as $p:=\rho_{\rm m}^2(\partial {\cal F}_0/\partial \rho_{\rm m})_T$. The internal dynamics of 
the director field is described by the Frank deformation potential ${\cal F}_{\rm d}$ which is defined in \cite{degennes}, for the simpler liquid crystal with group symmetry ($\infty/mm$), by 
\begin{align}    
{\cal F}_{\rm d}={}&\frac{1}{2}K_1({\bm \nabla}\cdot{\bm n})^2+\frac{1}{2}K_2({\bm n}\cdot{\bm \nabla}\times{\bm n})^2\nonumber\\
{}&+\frac{1}{2}K_3({\bm n}\times{\bm \nabla}\times {\bm n})^2.\label{fddd}
\end{align}
The three parameters $K_1$, $K_2$ and $K_3$ are known as Frank's elastic constants (elastic moduli), which are all independent from
each other and also positive. One usually calls $K_1$ splay, $K_2$ twist, and 
$K_3$ bend constants. The so-called saddle-splay boundary term is omitted, 
since it is a total derivative that does not contribute to the equations of 
motion. For a typical nematic crystal, one has $K_1 = 2.3\times 10^{-12}\,$N, 
$K_2 = 1.5\times 10^{-12}\,$N, $K_1 = 4.8\times 10^{-12}\,$N (see \cite{degennes,Virga}).

The cholesteric liquid crystals are characterized by an additional modulus $K_0$ 
and a constitutive constant $\tau$, related to the chirality of the
medium. As a result, the Frank potential (\ref{fddd}) is modified to
\begin{align}
{\cal F}_{\rm d}={}&K_0\tau({\bm n}\cdot{\bm \nabla}\times{\bm n}+ \tau) + \frac{1}{2}K_1({\bm \nabla}\cdot{\bm n})^2\nonumber\\
{}&+\frac{1}{2}K_2({\bm n}\cdot{\bm \nabla}\times{\bm n}+ \tau)^2+\frac{1}{2}K_3({\bm n}\times{\bm \nabla}\times {\bm n})^2.\label{fddd0}
\end{align}
We restrict ourselves to the case of the nematic crystals with $\tau = 0$,
although the generalization to the cholesteric crystals is straightforward.

The interaction free energy of the liquid crystal with an electric field ${\bm E}$ is represented by ${\cal F}_{\rm e}$. Generally, controlled in the field 
${\bm E}$, the electric free energy of the system \cite{LL1} reads ${\cal F}_{\rm e}={}-\int{\bm P}\cdot d{\bm E}$. Using (\ref{Pa}) and the expression for the permittivity tensor in the comoving frame (\ref{epij}), we explicitly obtain
\begin{align}
{\cal F}_{\rm e}=-\frac{1}{2}\varepsilon_0(\varepsilon_{\perp} - 1){\bm E}^2-\frac{1}{2}\varepsilon_0\Delta\varepsilon(\bm{n}\cdot\bm{E})^2.\label{fe3}
\end{align}
Analogously \cite{LL2,dejeu}, we have ${\cal F}_{\rm m} = -\int \bm{M}\cdot d\bm{B}$ for the magnetic free energy controlled in the field ${\bm B}$, which 
yields 
\begin{align}
{\cal F}_{\rm m}={}&-{\frac{1}{2\mu_0}}(1 - \mu^{-1}_{\perp}){\bm B}^2+\frac{1}{2\mu_0}\Delta\mu^{-1}(\bm{n}\cdot\bm{B})^2.\label{fm3}
\end{align}

Then, the nonrelativistic Lagrangian of the nematic liquid crystal is constructed as the difference ${\cal L}_{\rm nr} = {\cal K} - {\cal F}$ of the kinetic energy density ${\cal K}$ and the ``potential'' free energy density ${\cal F}$. Accordingly, we have
\begin{align}
{\cal L}_{\rm nr}={}&\rho_{\rm m}\frac{{\bm v}^2}{2}+\rho_{\rm m} {\cal J} \frac{{\bm \omega}^2}{2}-{\cal F}_0(\rho_{\rm m},T)-\,\frac{1}{2}K_1(\partial_an^a)^2\nonumber\\&{}-\frac{1}{2}K_2(\epsilon^{abc}n_a\partial_bn_c)^2-\frac{1}{2}K_3(\epsilon_{abc}n^b\epsilon^{cde}\partial_dn_e)^2\nonumber\\{}&+\,\frac{1}{2}\varepsilon_0\varepsilon_{\perp}{\bm E}^2+\frac{1}{2}\varepsilon_0\Delta\varepsilon(\bm{n}\cdot\bm{E})^2\nonumber\\
{}&-\frac{1}{2\mu_0}\mu^{-1}_{\perp}{\bm B^2}-\frac{1}{2\mu_0}\Delta\mu^{-1}(\bm{n}\cdot\bm{B})^2,\label{lenr}
\end{align}
where we added the energy density of the pure electromagnetic field $\varepsilon_0{\bm E}^2/2-\bm B^2/2\mu_0$. This is necessary to describe the electromagnetic field as a dynamical part of the system and to guarantee the correct limit of the energy density in free space.

The total non-relativistic liquid crystal Lagrangian (\ref{lenr}) can
be conveniently recast into the sum of the matter part ${\cal L}^{\rm m}_{\rm nr}(\rho_{\rm m},T,v^a,n^a,\partial_bn^a)$ and the electromagnetic part ${\cal L}^{\rm em}_{\rm nr}(n^a,E_a,B^a)$, 
\begin{align}
{\cal L}_{\rm nr}={\cal L}^{\rm m}_{\rm nr}+{\cal L}^{\rm em}_{\rm nr},
\end{align}
where
\begin{align}
{\cal L}^{\rm m}_{\rm nr}={}&\rho_{\rm m}\frac{{\bm v}^2}{2}+\rho_{\rm m} {\cal J} \frac{{\bm \omega}^2}{2}-{\cal F}_0(\rho_{\rm m},T)-\frac{1}{2}K_1(\partial_an^a)^2\nonumber\\{}&-\frac{1}{2}K_2(\epsilon^{abc}n_a\partial_bn_c)^2-\frac{1}{2}K_3(\epsilon_{abc}n^b\epsilon^{cde}\partial_dn_e)^2,\label{lenr3}\\
{\cal L}^{\rm em}_{\rm nr}={}&\frac{1}{2}\varepsilon_0\varepsilon_{\perp}{\bm E}^2+\frac{1}{2}\varepsilon_0\Delta\varepsilon(\bm{n}\cdot\bm{E})^2\nonumber\\{}&-\frac{1}{2}\mu_0^{-1}\mu^{-1}_{\perp}{\bm B^2}-\frac{1}{2}\mu_0^{-1}\Delta\mu^{-1}(\bm{n}\cdot\bm{B})^2.\label{lenrem}
\end{align}
The non-relativistic variational theory based on the Lagrangian (\ref{lenr}) was developed in \cite{lisin1,lisin2}. It used the Lagrange approach and was formulated in terms of $(\bm{E},\bm{H})$ instead of the fields $(\bm{E},\bm{B})$.
Dynamics of the total system is determined by the action integral $I_{\rm nr} =
\int dt\,d^3x\,{\cal L}_{\rm nr}$. 

\section{Relativistic liquid crystal Lagrangian}

The formal theory of liquid crystals has been studied only in the non-relativistic domain, as can be seen in \cite{degennes,chandra,dejeu,blinov,demus,scharf,Virga,stewart,erik5,erik6,leslie1,leslie2,lisin1,lisin2}. Here we develop a truly relativistic model for these systems, generalizing the non-relativistic 3-dimensional objects and operations to the corresponding 4-dimensional notions. In contrast to \cite{lisin1,lisin2}, we work in the Euler approach which is more convenient for field-theoretical applications. 

First, we notice that the 3-velocity field of the liquid crystal
$\bm{v}$ is proportional to the spatial part of the 4-velocity $u^i=(\gamma,\gamma {\bm v})$, which by definition always satisfies the condition
\begin{align}
u^iu_i=c^2>0.\label{connn2}
\end{align}
Thus, $u^i$ is a timelike 4-vector field. Here $\gamma = 1/\sqrt{1-v^2/c^2}$ is the usual Lorentz factor. Analogously, we can define
the \textit{director 4-vector} $N^i$ as the relativistic covariant generalization of the director $n^a$. When the medium is at rest,
$N^i$ should reduce to $n^a$; that is,
\begin{align}
{\stackrel {\circ}N}{}^i=(0,\bm{n}).\label{defni}
\end{align}
We now recall that $\bm{n}$, by definition, has a unit length. This together with (\ref{defni}) imposes the scalar condition
\begin{align}
N^iN_i=-1<0,\label{connn3}
\end{align}
which should be fulfilled in all reference frames. In other words, $N^i$ is a spacelike 4-vector. In the rest frame, ${\stackrel {\circ}u}{}^i = (1,\bm{0})$, and together with (\ref{defni}), we have ${\stackrel {\circ}u}{}_i{\stackrel {\circ}N}{}^i=0$. Since this is a scalar condition, it must be valid in all reference frames as well: 
\begin{align}
N^{i}u_{i}=0.\label{connn1}
\end{align}

In addition, it is necessary to define a relativistic generalization of the 3-dimensional Levi-Civita symbol in order to consistently express the ``cross products'' in (\ref{lenr}). Let us introduce
\begin{equation}
\epsilon_{ijk}:=\eta_{ijkl}\frac{u^{l}}{c},\label{eps}
\end{equation}
with the 4-dimensional Levi-Civita tensor defined such that $\eta_{0123}:=\sqrt{-g}=c$ and thus in the rest frame its spatial components reduce to the usual 3-dimensional Levi-Civita symbol ${\stackrel {\circ}\epsilon}{}^{abc}$, with ${\stackrel {\circ}\epsilon}{}^{123}=1$. Using this object, we immediately define the angular 4-velocity of the director
by
\begin{equation}
\omega^{i}:=\epsilon^{ijk}N_{j}\dot{N}_{k},\label{omega}
\end{equation}
where the convective ``time'' derivative is naturally
\begin{align}
\dot{N}^{i}=u^{j}\partial_{j}N^{i}.
\end{align}

The relativistic variational theory of an ideal fluid with structureless material elements is a well developed subject \cite{taub1,schutz,ray72,seliger,obukhov1}. 
The generalization to the ideal fluid with classical spin (modeled after the Dirac particles) was done in \cite{spinfluid1,spinfluid2}; see also the references therein.

A liquid crystal medium represents another example of a fluid with microstructure, represented by the director field attached to each element of the fluid. 
Here we develop a relativistic variational model for this system by combining the variational
model described in \cite{obukhov1} (for the kinetic translation energy and the internal energy), with the four-dimensional generalization of the liquid crystal elastic terms in (\ref{lenr3}):
\begin{align}
{\cal L}^{\rm m}={}&-\nu u^{i}\partial_{i}\Lambda_1+\Lambda_2 u^{i}\partial_{i}s+\Lambda_3 u^{i}\partial_{i}X-\frac{1}{2}J\nu\omega^{i}\omega_{i}\nonumber\\{}&-\rho(\nu,s)-\frac{1}{2}K_1\left(\partial_{i}N^{i}\right)^2-\frac{1}{2}K_2\left(\epsilon^{ijk}N_{i}\partial_{j}N_{k}\right)^2\nonumber\\{}&+\frac{1}{2}K_3\left(\epsilon_{ijk}N^{j}\epsilon^{kln}\partial_{l}N_{n}\right)^2+\Lambda_0(u^{i}u_{i}-c^2)\nonumber\\{}&+\Lambda_4(N^{i}N_{i}+1)+\Lambda_5 u^{i}N_{i}.\label{Lfluid}
\end{align}
Here $\nu$ is the particle number density of the liquid crystal, $J$ is the moment of inertia of one element [related to $\cal J$ in (\ref{kineticnonrel}) by means of ${\cal J}\rho_{\rm m}=J\nu$], $\rho(\nu,s)$ is the internal energy density of the relativistic fluid, $s$ is the entropy density, $X$ is the identity (Lin) coordinate, and $\Lambda_{I}$, with $I=0,\dots,5$ are Lagrange multipliers. By imposing these $\Lambda$'s we ensure the fulfillment of the conditions (\ref{connn2}), (\ref{connn3}), and (\ref{connn1}) throughout all the dynamics of the nematic liquid crystal, in addition to the particle number continuity equation,
\begin{align}
\partial_{i}(\nu u^{i})={}&0,
\end{align}
and the conservation of entropy and identity of particles along each streamline of the fluid:
\begin{align}
u^{i}\partial_{i}s={}&0,\\
u^{i}\partial_{i}X={}&0.
\end{align}
A different sign of the $K_3$ term in (\ref{Lfluid}), as compared to (\ref{lenr3}), is explained by the fact that the 4-vector $\epsilon_{ijk}N^{j}\epsilon^{kln}\partial_{l}N_{n}$ is spacelike, hence the square of its 4-length is negative. Notice that, as is usual in relativistic fluid models, we choose $s$ and $\nu$ as the indepedent thermodynamic quantities, instead of $T$ and $\rho_{\rm m}$ as in the non-relativistic case. This choice is purely convencional, since all the other thermodynamic quantities can be derived from them. 

The dynamics of the relativistic system is governed by the action $I =(1/c)
\int \sqrt{-g}\,d^4x\,{\cal L}^{\rm m}$. One may wonder how the nonrelativistic 
translational Lagrangian can be recovered from the relativistic Lagrangian.
As a first step, we represent the internal energy density as the sum $\rho = 
\rho_{\rm m}c^2 + {\cal F}_0$ of the ``rest-mass'' density and the hydrodynamic
energy density. Consider now an arbitrary volume element which reads 
$\sqrt{-g}\,d^4x = c\, dV_0\, d\tau$ in the comoving reference frame with the 3-volume
$dV_0$ and the proper time $\tau$. The next step is to notice that the rest
mass of a fluid's element $dm_0 = \rho_{\rm m}dV_0$ is the same in all frames and 
the invariant volume element $dV_0\, d\tau = dV dt$ [with $d\tau = \sqrt{1 - 
{\bm v}^2/c^2}dt \approx (1 - {\bm v}^2/2c^2)dt$] in the reference frame where the fluid's element $dV$ has velocity ${\bm v}$. As a result, in the nonrelativistic
limit we indeed recover the translational part of the Lagrangian: $(1/c)
\int \sqrt{-g}d^4x\,(-\rho)\approx \int dt dV\left(\rho_{\rm m}{\bm v}^2/2 - {\cal F}_0\right)$. A more detailed discussion can be found in \cite{schutz2}, for example.

\section{Lagrangian for the electromagnetic field interacting with the liquid crystal medium}

In order to describe the interaction of the electromagnetic field with matter in an explicitly covariant manner, we make use of the standard electromagnetic Lagrangian given by
\begin{align}
{\cal L}^{\rm em}=-\,\frac{1}{4}H^{ij}F_{ij}.\label{emlagrangian}
\end{align}
It yields the macroscopic Maxwell equations as Euler-Lagrange equations without sources (we assume that the fluid elements are not electrically charged):
\begin{align}
\partial_jH^{ij}=0.\label{macromax}
\end{align}
Here the electromagnetic strength tensor $F_{ij}$ is expressed in terms of the electromagnetic 4-potential $A_i$, as usual by $F_{ij}:=\partial_iA_j-\partial_jA_i$ and $H^{ij}$ is the covariant electromagnetic excitation tensor, \cite{obukhov1,birk}.

The constitutive relations for any linear, non-dissipative and non-dispersive medium can be expressed in general covariant form by
\begin{align}
H^{ij}=\frac{1}{2}\chi^{ijkl}F_{kl},\label{defhij}
\end{align}
where $\chi^{ijkl}$ is the so-called constitutive tensor with the symmetries
\begin{align}
\chi^{ijkl}=-\chi^{jikl}=-\chi^{ijlk}=\chi^{klij}.
\end{align}
Therefore, it has $21$ independent components, in general. If we insert (\ref{defhij}) into (\ref{emlagrangian}), we obtain
\begin{align}
{\cal L}^{\rm em}=-\,\frac{1}{8}\chi^{ijkl}F_{ij}F_{kl}.\label{emlagrangianchi}
\end{align}
Historically, the general constitutive relation (\ref{defhij}) was first formulated by Bateman \cite{Bateman}, Tamm \cite{Tamm1,Tamm2,Tamm3}, and later in the modern notation by Post \cite{Post}.

As we commented in Sec. \ref{constirest}, the nematic liquid crystal is an example of a uniaxial dielectric and diamagnetic anisotropic medium and therefore we need to find an explicit covariant expression for the constitutive tensor $\chi^{ijkl}$ for 
a medium of this kind.

\subsection{Constitutive tensor in the comoving frame}

The components of the covariant constitutive relation (\ref{defhij}), must reproduce the expressions (\ref{epij}) and (\ref{invmuij}), in the rest frame of the medium. Therefore, given ${\stackrel {\circ}\varepsilon}{}^{ab}$ and ${\stackrel {\circ}\mu}{}^{-1}_{ab}$ in terms of the director $\bm{n}$ and the eigenvalues $\varepsilon_{\perp}$, $\varepsilon_{\|}$, $\mu^{-1}_{\perp}$, and $\mu^{-1}_{\|}$, the non-vanishing components of ${\stackrel {\circ}\chi}{}^{ijkl}$ must explicitly read
\begin{align}
{\stackrel {\circ}\chi}{}^{0ab0}={}&{\stackrel\circ\varepsilon}{}^{ab} 
          = \varepsilon_{\perp} \delta^{ab}+\Delta\varepsilon\ {}n^an^b,\label{chicom11}\\
{\stackrel {\circ}\chi}{}^{abcd}={}&\epsilon^{abf}\epsilon^{cdg}{\stackrel {\circ}\mu}{}^{-1}_{fg} \nonumber\\
          ={}&\mu^{-1}_{\|}\left(\delta^{ac}\delta^{bd}-\delta^{ad}\delta^{bc}\right)\nonumber\\ 
           {}&-\Delta\mu^{-1}\left(\delta^{ac}n^bn^d-\delta^{ad}n^bn^c\right.\nonumber\\
           {}&\hspace{1.5cm}\left.+\delta^{bd}n^an^c -\delta^{bc}n^an^d\right).\label{chicom22}
\end{align}

It is worthwhile to notice that this constitutive tensor characterizes all non-magnetoelectric anisotropic media with dielectric and diamagnetic uniaxial properties in the direction of the optical axis ${\bm n}$. In the special case of nematic liquid crystals, the optical axis coincides with the director field.

\subsection{Dispersion relations and factorization of the Fresnel equation}

If we look for wave solutions to the macroscopic Maxwell's equations (\ref{macromax}) in a source-free and homogeneous medium (described by $\chi^{ijkl}$), then the general dispersion relation is determined in covariant form by the fourth order Fresnel equation for the 4-wave covector $k_{i}$:
\begin{align}
{\cal G}^{ijkl}k_{i}k_{j}k_{k}k_{l}=0. \label{fresnel}
\end{align}
Here ${\cal G}^{ijkl}$ is the Tamm-Rubilar tensor \cite{obukhovfukuirubilar}, given by
\begin{align}
{\cal G}^{ijkl}:=\frac{1}{4!c^2}\eta_{mnpq}\eta_{rstu}\chi^{mnr(i}\chi^{j|ps|k}\chi^{l)qtu}. \label{TR}
\end{align}

We can use the non-vanishing components of the constitutive tensor (\ref{chicom11}) and (\ref{chicom22}) in (\ref{TR}) and (\ref{fresnel}) and thereby verify the factorization of the 4-rth order Fresnel wave surface into a product of two light cones, determined by two optical metrics in the rest frame of the medium:
\begin{align}
{\stackrel {\circ}{\cal G}}{}^{ijkl}k_{i}k_{j}k_{k}k_{l} = ({\stackrel {\circ}\gamma}{}^{ij}_{\rm e}\,k_{i}k_{j})({\stackrel {\circ}\gamma}{}^{kl}_{\rm m}\,k_{k}k_{l})=0.
\end{align}
Here the light cones read explicitly
\begin{eqnarray}
{\stackrel {\circ}\gamma}{}^{ij}_{\rm e}\,k_{i}k_{j} &=& n^2k_0^2-\alpha_{\rm e}{\bm k}^2+(\alpha_{\rm e}-1)({\bm n}\cdot{\bm k})^2 ,\label{gekk}\\
{\stackrel {\circ}\gamma}{}^{ij}_{\rm m}\,k_{i}k_{j} &=& n^2k_0^2-\alpha_{\rm m}{\bm k}^2+(\alpha_{\rm m}-1)({\bm n}\cdot{\bm k})^2,\label{gmkk}
\end{eqnarray}
with the refractive index
\begin{align}
n^2:=\mu_{\perp}\varepsilon_{\perp}.
\end{align}
The parameters $\alpha_{\rm e}$ and $\alpha_{\rm m}$ quantify the proportion of dielectric and diamagnetic uniaxial anisotropy in the medium, and read
\begin{align}
\alpha_{\rm e}:=\frac{\varepsilon_{\perp}}{\varepsilon_{\|}}, \qquad
\alpha_{\rm m}:=\frac{\mu_{\perp}}{\mu_{\|}}.
\end{align}

One of the reduced quadratic Fresnel dispersion relations, 
${\stackrel {\circ}\gamma}{}^{ij}_{\rm e}\,k_{i}k_{j}=0$, implies that if ${\bm k}$ is parallel to 
${\bm n}$, then $n^2k_0^2-k^2=0$, which means that in this case light propagates 
with the expected effective refraction index of the ordinary ray: $n$. On the other hand, if 
${\bm k}$ is orthogonal to ${\bm n}$ then the dispersion relation reduces to 
$n^2k_0^2-\alpha_{\rm e}k^2=0$. This means that light propagates with an effective refraction 
index $n_{\rm e}$, given by $n_{\rm e}^2=n^2/\alpha_{\rm e}=\varepsilon_\parallel\mu_\perp$. We may call this 
the \emph{dielectric extraordinary ray}. Similarly, the second Fresnel dispersion relation, 
${\stackrel {\circ}\gamma}{}^{ij}_{\rm m}k_{i}k_{j}=0$, leads to a normal ordinary ray refraction 
index $n$ for waves with wave vector ${\bm k}$ parallel to the optical axis ${\bm n}$. For ${\bm k}\perp{\bm n}$, it implies that light propagates with a refraction index $n_{\rm m}$ , with $n^2_{\rm m}=n^2/\alpha_{\rm m}=\varepsilon_\perp\mu_\parallel$, 
corresponding to the \emph{diamagnetic extraordinary ray}.

\subsection{Covariant description}\label{consttensor}

Since both (\ref{gekk}) and (\ref{gmkk}) must be covariant equations, it is not difficult to 
derive the general expressions of the two optical metrics in a frame where the medium moves with 
an arbitrary 4-velocity $u^i=(\gamma,\gamma\bm{v})$:
\begin{align}
\gamma^{ij}_{\rm e}={}&\alpha_{\rm e}\, g^{ij}+\frac{(n^2-\alpha_{\rm e})}{c^2}u^{i}u^{j}+(\alpha_{\rm e}-1)N^{i}N^{j},\label{dielectric}\\
\gamma^{ij}_{\rm m}={}&\alpha_{\rm m}\, g^{ij}+\frac{(n^2-\alpha_{\rm m})}{c^2}u^{i}u^{j}+(\alpha_{\rm m}-1)N^{i}N^{j}.\label{diamagnetic}
\end{align}
One can then verify that the covariant form of the constitutive tensor, which takes into account all the necessary symmetries and which reduces to (\ref{chicom11}) and (\ref{chicom22}) in the rest frame of the medium, is given by
\begin{align}
\chi^{ijkl}={}&\frac{1}{\mu_0}\left(\mu^{-1}_{\perp}+\Delta\mu^{-1}\right)\left(g^{ik}g^{jl}-g^{il}g^{jk}\right)\nonumber\\+{}&\frac{1}{\mu_0c^2}\left[\frac{\left(n^2-1\right)}{\mu_{\perp}}-\Delta\mu^{-1}\right]\left(g^{ik} u^{j} u^{l}-g^{il} u^{j} u^{k}\right.\nonumber\\
{}&\hspace{4cm}\left.+g^{jl} u^{i} u^{k}-g^{jk} u^{i} u^{l}\right)\nonumber\\
+{}&\frac{1}{\mu_0}\Delta\mu^{-1}\left(g^{ik} N^{j} N^{l}-g^{il} N^{j} N^{k}\right.\nonumber\\
{}&\hspace{1.4cm}\left.+g^{jl} N^{i} N^{k}-g^{jk} N^{i} N^{l}\right)\nonumber\\
-{}&\frac{1}{\mu_0c^2}\left(\Delta\varepsilon+\Delta\mu^{-1}\right)\left(u^{i}u^{k}N^{j}N^{l}-u^{i}u^{l}N^{j}N^{k}\right.\nonumber\\
{}&\hspace{1.6cm}\left.+u^{j}u^{l}N^{i}N^{k}-u^{j}u^{k}N^{i}N^{l}\right).\label{explicit}
\end{align}

Generalizing the result obtained by Balakin and Zimdahl in \cite{balakin}, the constitutive tensor can also be expressed in terms of the two optical metrics (\ref{dielectric}) and (\ref{diamagnetic}), by
\begin{align}
\chi^{ijkl}={}&\frac{1}{\alpha_{\rm e}\,\mu_0\mu_{\perp}}\left(\gamma^{ik}_{\rm e}\gamma^{jl}_{\rm e}-\gamma^{il}_{\rm e}\gamma^{jk}_{\rm e}\right)\nonumber\\
{}&+\frac{1}{(\alpha_{\rm m}-\alpha_{\rm e})\mu_0\mu_{\perp}}\left(\Delta\gamma^{ik}\Delta\gamma^{jl}-\Delta\gamma^{il}\Delta\gamma^{jk}\right),\label{chiani}
\end{align}
where $\Delta\gamma^{ij}$ is the difference of the optical metrics, 
\begin{align}
\Delta\gamma^{ij}:={}&\gamma^{ij}_{\rm e}-\gamma^{ij}_{\rm m}.\label{difgam}
\end{align}

\subsection{The projector $\pi^{ij}$ and the isotropic limit}

Let us define the projector $\pi^{ij}$ to the 2-dimensional space orthogonal to $N^{i}$ and $u^{i}$,
\begin{align}
\pi^{ij}:=g^{ij}-\frac{1}{c^2}u^{i}u^{j}+N^{i}N^{j}.\label{pi}
\end{align}
It obviously has the properties $\pi^{i}{}_{j}\pi^{j}{}_{k}=\pi^{i}{}_{k}$ and $\det{(\pi^{ij})}= 0$.
With this object we can write the optical metrics in a more compact form:
\begin{align}
\gamma^{ij}_{\rm e}={}&\gamma^{ij}+(\alpha_{\rm e}-1)\pi^{ij},\label{dielectric2}\\
\gamma^{ij}_{\rm m}={}&\gamma^{ij}+(\alpha_{\rm m}-1)\pi^{ij},\label{diamagnetic2}
\end{align}
where $\gamma^{ij}$ is the usual optical Gordon metric of an isotropic medium \cite{obukhov1,wgordon},
\begin{align}
\gamma^{ij}=g^{ij}+\frac{(n^2-1)}{c^2}u^{i}u^{j}.\label{gordon}
\end{align}
Also the difference of optical metrics (\ref{difgam}) can be expressed in terms of $\pi^{ij}$,
\begin{align}
\Delta\gamma^{ij}=(\alpha_{\rm e}-\alpha_{\rm m})\pi^{ij}.
\end{align}
Therefore, the constitutive tensor in (\ref{chiani}) can be recast into the form
\begin{align}
\chi^{ijkl}={}&\frac{1}{\mu_0\mu_{\perp}}\left[\frac{1}{\alpha_{\rm e}}\left(\gamma^{ik}_{\rm e}\gamma^{jl}_{\rm e}-\gamma^{il}_{\rm e}\gamma^{jk}_{\rm e}\right)\right.\nonumber\\
{}&\hspace{1.1cm}\left.-(\alpha_{\rm e}-\alpha_{\rm m})\left(\pi^{ik}\pi^{jl}-\pi^{il}\pi^{jk}\right)\right].\label{chiani2}
\end{align}
{}From (\ref{chiani2}), we can easily check that our $\chi^{ijkl}$ reduces to the well-known expression for the isotropic case in terms of the Gordon metric (\ref{gordon}), when 
\begin{align}
\varepsilon_{\|}\rightarrow\varepsilon_{\perp}=\varepsilon &\quad\Leftrightarrow\quad \alpha_{\rm e}\rightarrow 1,\\
\mu_{\|}\rightarrow\mu_{\perp}=\mu &\quad\Leftrightarrow\quad \alpha_{\rm m}\rightarrow 1.
\end{align}
Then both optical metrics in (\ref{dielectric2}) and (\ref{diamagnetic2}) become the single Gordon metric, and the constitutive tensor reduces to
\begin{align}
\chi^{ijkl}\rightarrow \chi^{ijkl}_{\rm iso}:=\frac{1}{\mu_0\mu}\left(\gamma^{ik}\gamma^{jl}-\gamma^{il}\gamma^{jk}\right).\label{chiiso}
\end{align}

\subsection{Explicit expression for the electromagnetic Lagrangian}

With the help of the constitutive tensor (\ref{explicit}), we can use (\ref{defhij}) to obtain an explicit expression for the electromagnetic excitation $H^{ij}$:
\begin{align}
H^{kl}={}&\frac{1}{\mu_0}\left(\mu^{-1}_{\perp} + \Delta\mu^{-1}\right)F^{kl}+\frac{2}{\mu_0}\Delta\mu^{-1}\,F^{[k}{}_n N^{l]}N^n\nonumber\\
{}&+ {\frac {2}{\mu_0c^2}}\left(\varepsilon_{\perp} - \mu^{-1}_{\perp} - \Delta\mu^{-1}\right)F^{[k}{}_n u^{l]}u^n\nonumber\\
{}&- {\frac {2}{\mu_0c^2}}\left(\Delta\varepsilon + \Delta\mu^{-1}\right)N^{[k}u^{l]}\,F_{pq}N^p u^q.\label{Hab}
\end{align}
From here on, we assume that the permittivity and permeability can be functions of the particle density, $\mu_{\perp}=\mu_\bot(\nu)$, $\varepsilon_{\perp}=\varepsilon_{\bot}(\nu)$, and $\Delta\mu^{-1}=\Delta\mu^{-1}(\nu)$, in general. Then, from (\ref{emlagrangian}) and (\ref{Hab}), it is straightforward to compute an explicit expression for the electromagnetic Lagrangian ${\cal L}^{\rm em}={\cal L}^{\rm em}(\nu,u^i,N^i,F_{ij})$, which reads
\begin{align}
{\cal L}^{\rm em}{}&= -\,{\frac 1{4\mu_0}}(\mu^{-1}_{\perp}+\Delta\mu^{-1})F_{ij}F^{ij} - {\frac 1{2\mu_0}}\Delta\mu^{-1}\,(F_{kl} N^l)^2\nonumber\\
{}&- {\frac 1{2\mu_0c^2}}\left(\varepsilon_{\perp} - \mu^{-1}_{\perp} - \Delta\mu^{-1}\right)(F_{kl}u^l)^2\nonumber\\
{}&+{\frac 1{2\mu_0c^2}}\left(\Delta\varepsilon + \Delta\mu^{-1}\right)
(F_{pq}N^p u^q)^2.\label{Lem1}
\end{align}

For completeness, we also give an alternative derivation of (\ref{Lem1}) based directly on the non-relativistic Lagrangian (\ref{lenrem}). In the four-dimensional relativistic framework, the electric $\bm{E}$ and magnetic $\bm{B}$ fields are substituted with the 4-vectors of electric field ${\cal E}_{i}$ and magnetic field ${\cal B}^{i}$, defined as
\begin{align}
{\cal E}_{i}:={}&F_{ij}u^{j},\label{4e}\\
{\cal B}^{i}:={}&\frac{1}{2c}\eta^{ijkl}F_{jk}u_{l}.\label{4b}
\end{align}
Then the electromagnetic Lagrangian (\ref{emlagrangian}) can be alternatively written in a ``3-D like'' form,
\begin{align}
{\cal L}^{\rm em}=\frac{1}{2}\left(\varepsilon_0\varepsilon^{ij}{\cal E}_{i}{\cal E}_{j}-\mu_0^{-1}\mu^{-1}_{ij}{\cal B}^{i}{\cal B}^{j}\right),\label{4lem}
\end{align}
where $\varepsilon^{ij}$ is the 4-permittivity tensor and $\mu^{-1}_{ij}$ the inverse of the 4-permeability tensor, given by
\begin{align}
\varepsilon^{ij}:={}&-\varepsilon_{\perp}g^{ij}+\Delta\varepsilon N^{i}N^{j},\label{4epsilon}\\
\mu^{-1}_{ij}:={}&-\mu^{-1}_{\perp}g_{ij}+\Delta\mu^{-1} N_{i}N_{j}.\label{4mu}
\end{align}
Notice that the spatial components of these expressions consistently reduce to (\ref{epij}) and (\ref{invmuij}). 
Inserting (\ref{4epsilon}) and (\ref{4mu}) into (\ref{4lem}), we can separate the contributions of the electromagnetic Lagrangian in ``isotropic'' and ``anisotropic'' parts, that is,
\begin{align}
{\cal L}^{\rm em}={\cal L}^{\rm iso}+{\cal L}^{\rm ani},\label{Lem}
\end{align}
with
\begin{align}
{\cal L}^{\rm iso}={}&-\frac{1}{2}\left[\varepsilon_0\varepsilon_{\perp}{\cal E}^{i}{\cal E}_{i}-\mu_0^{-1}\mu^{-1}_{\perp}{\cal B}^{i}{\cal B}_{i}\right],\label{liso}\\
{\cal L}^{\rm ani}={}&\frac{1}{2}\left[\varepsilon_0\Delta\varepsilon ({\cal E}_{i}N^{i})^2-\mu_0^{-1}\Delta\mu^{-1} ({\cal B}_{i}N^{i})^2\right].\label{lani}
\end{align}
As we see, the structure of (\ref{Lem}), (\ref{liso}), and (\ref{lani}) follows exactly the structure of the non-relativistic Lagrangian (\ref{lenrem}). By inserting (\ref{4e}) and (\ref{4b}) into (\ref{liso}), we can write the isotropic electromagnetic Lagrangian ${\cal L}^{\rm iso}$ in terms of $F^{ij}$:
\begin{align}
{\cal L}^{\rm iso}={}&-\frac{1}{4\mu_0\mu_{\perp}}g^{ik}g^{jl}F_{ij}F_{kl}\nonumber\\
{}&-\frac{(n^2-1)}{2\mu_0\mu_{\perp}c^2}g^{ik}u^{j}u^{l}F_{ij}F_{kl}\\
                  ={}&-\frac{1}{4\mu_0\mu_{\perp}}\gamma^{ik}\gamma^{jl}F_{ij}F_{kl}\\
                  ={}&-\frac{1}{8}\chi^{ijkl}_{\rm iso}F_{ij}F_{kl},
\end{align}
where $\chi^{ijkl}_{\rm iso}$ is the constitutive tensor for the isotropic case given in (\ref{chiiso}) and written in terms of the Gordon metric (\ref{gordon}). In this case $n^2=\varepsilon_{\perp}\mu_{\perp}$ as expected.

Now we do the same for the anisotropic part of the electromagnetic Lagrangian (\ref{lani}). First, using the definitions (\ref{4e}) and (\ref{4b}), we explicitly have the following expressions:
\begin{align}
\left(N^{i}{\cal B}_{i}\right)^2={}&\frac{1}{2}F^{ij}F_{ij}+(F_{ij}N^{j})^2-\frac{1}{c^2}(F_{ij}u^{j})^2\nonumber\\
{}&-\frac{1}{c^2}(F_{ij}N^{i}u^{j})^2,\label{nB}\\
\left(N^{i}{\cal E}_{i}\right)^2={}&(F_{ij}N^{i}u^{j})^2.\label{nE}
\end{align}
Then, using (\ref{nB}) and (\ref{nE}) in (\ref{lani}), we derive the explicit expression for the anisotropic Lagrangian,
\begin{align}
{\cal L}^{\rm ani}={}&-\frac{1}{4\mu_0}\Delta\mu^{-1}g^{ik}g^{jl}F_{ij}F_{kl}\nonumber\\
{}&-\frac{1}{2\mu_0}\Delta\mu^{-1}g^{ik}N^{j}N^{l}F_{ij}F_{kl}\nonumber\\
{}&+\frac{1}{2\mu_0 c^2}\Delta\mu^{-1} g^{ik}u^{j}u^{l}F_{ij}F_{kl}\nonumber\\
{}&+\frac{1}{2\mu_0 c^2}(\Delta\varepsilon+\Delta\mu^{-1})N^{i}N^{k}u^{j}u^{l}F_{ij}F_{kl}\\
                ={}&-\frac{1}{8}\chi^{ijkl}_{\rm ani}F_{ij}F_{kl},
\end{align}
with
\begin{align}
\chi^{ijkl}_{\rm ani}:={}&\mu_0^{-1}\Delta\mu^{-1}\left(g^{ik}g^{jl}-g^{il}g^{jk}\right)\nonumber\\
{}&-\frac{1}{\mu_0c^2}\Delta\mu^{-1}\left(g^{ik} u^{j} u^{l}-g^{il} u^{j} u^{k}\right.\nonumber\\
{}&\hspace{2.1cm}\left.+g^{jl} u^{i} u^{k}-g^{jk} u^{i} u^{l}\right)\nonumber\\
{}&+\mu_0^{-1}\Delta\mu^{-1}\left(g^{ik} N^{j} N^{l}-g^{il} N^{j} N^{k}\right.\nonumber\\
{}&\hspace{2.1cm}+\left.g^{jl} N^{i} N^{k}-g^{jk} N^{i} N^{l}\right)\nonumber\\
{}&-\frac{1}{\mu_0c^2}\left(\Delta\varepsilon+\Delta\mu^{-1}\right)\left(u^{i}u^{k}N^{j}N^{l}-u^{i}u^{l}N^{j}N^{k}\right.\nonumber\\
{}&\hspace{1.7cm}+\left.u^{j}u^{l}N^{i}N^{k}-u^{j}u^{k}N^{i}N^{l}\right).
\end{align}

Finally, the total constitutive tensor $\chi^{ijkl}$ in (\ref{explicit}), is formed by adding (\ref{chiani}) and (\ref{chiiso}), as can be easily checked,
\begin{align}
\chi^{ijkl}=\chi^{ijkl}_{\rm iso}(\varepsilon_{\perp},\mu_{\perp})+\chi^{ijkl}_{\rm ani}(\Delta\varepsilon,\Delta\mu^{-1}).
\end{align}
The resulting constitutive tensor is identical to (\ref{explicit}).

\section{Variation of the matter Lagrangian}\label{fieldeqs}

With a well defined Lagrangian (\ref{Lfluid}) for our relativistic model of a nematic liquid crystal, we can now derive the field equations of the system. For this task, we write the matter Lagrangian ${\cal L}^{\rm m}={\cal L}^{\rm m}(\nu,u^{i},N^{i},\partial_jN^i,s,X,\Lambda_{I})$, $I=0,1,2,3,4,5$, as a sum of three terms,
\begin{align}\label{Lm}
{\cal L}^{\rm m}={\cal L}^{\rm k} + {\cal L}^{\rm p} + {\cal L}^{\Lambda},
\end{align}
where
\begin{align}
{\cal L}^{\rm k}:= -\,{\frac 1 2}\,J\nu\,\omega^i\omega_i \label{Lk}
\end{align}
is the kinetic Lagrangian and
\begin{align}
{\cal L}^{\rm p}:= -\rho(\nu,s) - {\cal V}\label{Lp}
\end{align}
is the Lagrangian of the potential energy, with the Frank deformation potential given by
\begin{align}
{\cal V}:={}& {\frac 12}K_1\left(\partial_{i}N^{i}\right)^2 + {\frac 12}
K_2\left(\epsilon^{ijk}N_{i}\partial_{j}N_{k}\right)^2\nonumber\\
 {}&- {\frac 12}K_3\left(\epsilon_{ijk}N^{j}\epsilon^{kln}
\partial_{l}N_{n}\right)^2.\label{VF}
\end{align}
The Lagrangian part containing the constraints reads
\begin{align}
{\cal L}^{\Lambda}:={}&\Lambda_0(u^{i}u_{i}-c^2)-\nu u^{i}\partial_{i}
\Lambda_1+\Lambda_2 u^{i}\partial_{i}s\nonumber\\
{}&+\Lambda_3 u^{i}\partial_{i}X
+\Lambda_4(N^{i}N_{i}+1)+\Lambda_5 u^{i}N_{i}.\label{Ll}
\end{align}

\subsection{Kinetic term}

To begin with, we notice that making use of the definition (\ref{omega}), we
can recast the square of the angular velocity into
\begin{align}\label{omega2}
\omega^i\omega_i=P^i{}_j\,\dot{N}_i\dot{N}^j= 
\dot{N}^i\dot{N}_i-{\frac 1{c^2}}\left(u_i\dot{N}^i\right)^2,
\end{align}
where $P^i{}_j$ is the usual projector operator, perpendicular to the 4-velocity field:
\begin{align}
P^i{}_j:=\delta^i_j-\frac{1}{c^2}u^i u_j.\label{projectoruu}
\end{align}

Therefore, the kinetic Lagrangian (\ref{Lk}) depends only on $\nu, u^i$, and the derivatives
of the director field $\partial_k N^i$. The derivative with respect to
the latter reads
\begin{align}
{\frac {\partial{\cal L}^{\rm k}}{\partial\partial_k N^i}} =
- J\nu\,u^k\,P^j{}_i\,\dot{N}_j.\label{dLk_ddN}
\end{align}
As a result, we find the variational derivatives of the kinetic Lagrangian
with respect to its arguments
\begin{align}
{\frac {\delta{\cal L}^{\rm k}}{\delta\nu}} &= {\frac {\partial{\cal L}^{\rm k}}
{\partial\nu}} = -\,{\frac J2}\omega^i\omega_i,\label{dLk_nu}\\
{\frac {\delta{\cal L}^{\rm k}}{\delta u^i}} &= {\frac {\partial
{\cal L}^{\rm k}}{\partial u^i}} = J\nu\left(-P^j{}_k\partial_i
N^k + \dot{N}_i u^j/c^2\right)\dot{N}_j,\label{dLk_du}\\
{\frac {\delta{\cal L}^{\rm k}}{\delta N^i}} &= -\,\partial_k\left(
{\frac {\partial {\cal L}^{\rm k}}{\partial\partial_k N^i}}\right) =
\partial_k\left(J\nu\,u^k\,P^j{}_i\,\dot{N}_j\right).\label{dLk_dN}
\end{align}

\subsection{Potential term}

The potential Lagrangian (\ref{Lp}) depends on $\nu, s, u^i, N^i$, and 
the derivatives of the director field $\partial_k N^i$. The derivative 
with respect to the latter reads
\begin{align}
{\frac {\partial{\cal L}^{\rm p}}{\partial\partial_k N^i}} =
-\,{\frac {\partial{\cal V}}{\partial\partial_k N^i}}.\label{dLp_ddN}
\end{align}
We straightforwardly compute the variational derivatives
\begin{align}
{\frac {\delta{\cal L}^{\rm p}}{\delta\nu}} &= {\frac {\partial{\cal L}^{\rm p}}
{\partial\nu}} = -\,{\frac {\partial\rho}{\partial\nu}} = -\,{\frac 
{p + \rho}\nu},\label{dLp_dnu}\\
{\frac {\delta{\cal L}^{\rm p}}{\delta s}} &= {\frac {\partial{\cal L}^{\rm p}}
{\partial s}} = -\,{\frac {\partial\rho}{\partial s}}=-\,\nu T,\label{dLp_ds}\\
{\frac {\delta{\cal L}^{\rm p}}{\delta u^i}} &= {\frac {\partial
{\cal L}^{\rm p}}{\partial u^i}} = -\,{\frac {\partial{\cal V}}{\partial
u^i}},\label{dLp_du}\\
{\frac {\delta{\cal L}^{\rm p}}{\delta N^i}} &= -\,{\frac {\delta{\cal V}}
{\delta N^i}} = -\,{\frac {\partial{\cal V}}{\partial N^i}}
+ \,\partial_k\left({\frac {\partial {\cal V}}{\partial\partial_k N^i}}
\right).\label{dLp_dN}
\end{align}
In (\ref{dLp_dnu}) and (\ref{dLp_ds}) we used the thermodynamic (Gibbs) law
\begin{align}
Tds = d\left(\frac{\rho}{\nu}\right) + pd\left(\frac{1}{\nu}\right).\label{Gibbs}
\end{align}

\subsection{Constraint term}

Variation with respect to the Lagrange multipliers $\Lambda_{I}$ yields 
the complete set of constraints:
\begin{align}
u^i u_i ={}& c^2,\label{C0}\\
\partial_{i}(\nu u^{i})={}&0,\label{C1}\\
u^{i}\partial_{i}s={}&0,\label{C2}\\
u^{i}\partial_{i}X={}&0,\label{C3}\\
N^i N_i ={}& -\,1,\label{C4}\\
N^i u_i ={}& 0.\label{C5}
\end{align}
Additionally, the variations of the constraint Lagrangian with respect to the field variables read
\begin{align}
{\frac {\delta{\cal L}^{\Lambda}}{\delta\nu}}&={\frac{\partial{\cal L}^{\Lambda}}
{\partial\nu}} = -\,u^i\partial_i\Lambda_1,\label{dLl_dnu}\\
{\frac {\delta{\cal L}^{\Lambda}}{\delta s}}&=-\,\partial_i\left({\frac{\partial
{\cal L}^{\Lambda}}{\partial \partial_is}}\right) = -\,\partial_i(u^i\Lambda_2),
\label{dLl_ds}\\
{\frac {\delta{\cal L}^{\Lambda}}{\delta X}}&=-\,\partial_i\left({\frac{\partial
{\cal L}^{\Lambda}}{\partial \partial_iX}}\right) = -\,\partial_i(u^i\Lambda_3),
\label{dLl_dX}\\
{\frac {\delta{\cal L}^{\Lambda}}{\delta u^i}}& = {\frac{\partial
{\cal L}^{\Lambda}} {\partial u^i}} = 2\Lambda_0u_i - \nu\partial_i
\Lambda_1 + \Lambda_2\partial_i s\nonumber\\
&\hspace{1.6cm}+ \Lambda_3\partial_i X+\Lambda_5 N_i,\label{dLl_du}\\ \label{dLl_dN}
{\frac {\delta{\cal L}^{\Lambda}}{\delta N^i}}& = {\frac{\partial
{\cal L}^{\Lambda}} {\partial N^i}} =2\Lambda_4N_i + \Lambda_5u_i. 
\end{align}

\subsection{Field equations}

We are now in a position to write the field equations, collecting the
variations of the three terms in (\ref{Lm}). We can distinguish three groups 
of equations. The first group describes the constraints (\ref{C0})--(\ref{C5}).
The second group concerns the variables $s, X$ which do not enter the 
Lagrangian of the electromagnetic field. This yields 
\begin{align}
\partial_i(u^i\Lambda_2) + \nu T &= 0,\label{Es}\\
\partial_i(u^i\Lambda_3) &= 0.\label{EX}
\end{align}
The third group of equations is obtained from the variations with respect to
the essential field variables $\nu$, $u^i$, and $N^i$. We have
explicitly
\begin{align}
{\frac {\delta{\cal L}^{\rm m}}{\delta\nu}} =& -\,{\frac J2}\,\omega^2 
- {\frac {p + \rho}\nu} - u^i\partial_i\Lambda_1,\label{dLm_dnu}\\
{\frac {\delta{\cal L}^{\rm m}}{\delta u^i}} =& J\nu\left(-P^j{}_k
\partial_i N^k +\frac{1}{c^2}\dot{N}_i u^j\right)\dot{N}_j 
- {\frac {\partial{\cal V}}{\partial u^i}}\nonumber\\
& + 2\Lambda_0u_i - \nu\partial_i\Lambda_1 + \Lambda_2\partial_i s
+ \Lambda_3\partial_i X + \Lambda_5 N_i,\label{dLm_du}\\
{\frac {\delta{\cal L}^{\rm m}}{\delta N^i}} =& \partial_k\left(J\nu
\,u^k\,P^j{}_i\,\dot{N}_j\right) -\,{\frac {\delta{\cal V}}
{\delta N^i}} + 2\Lambda_4N_i + \Lambda_5u_i.\label{dLm_dN}
\end{align}
Contracting (\ref{dLm_dN}) with $u^i$ and $N^i$, we find the Lagrange
multipliers
\begin{align}\label{Lambda5}
\Lambda_5 &= {\frac 1{c^2}}u^i\left[{\frac {\delta{\cal L}^{\rm m}}
{\delta N^i}} - \partial_k\left(J\nu\,u^k\,P^j{}_i
\,\dot{N}_j\right) + {\frac {\delta{\cal V}}{\delta N^i}}\right],\\
2\Lambda_4 &= -\,N^i\left[{\frac {\delta{\cal L}^{\rm m}}{\delta N^i}} 
- \partial_k\left(J\nu\,u^k\,P^j{}_i\,\dot{N}_j\right) 
+ {\frac {\delta{\cal V}}{\delta N^i}}\right].\label{Lambda4}
\end{align}
Substituting these back into (\ref{dLm_dN}), we end with the field equation
for the director field:
\begin{align}\label{EqN}
\pi_i{}^j{\frac {\delta{\cal L}^{\rm m}}{\delta N^j}} = \pi_i{}^j\left[
\partial_k\left(J\nu\,u^k\,P^l{}_j\,\dot{N}_l\right) - {\frac {\delta{\cal V}}
{\delta N^j}}\right].
\end{align}
Here $\pi_i{}^j$ is the projector defined in (\ref{pi}). Note that we cannot yet
put equal zero the left-hand side of (\ref{EqN}); it will be evaluated later
from the variation of the electromagnetic part (\ref{Lem1}) of the total Lagrangian. 

Contracting (\ref{dLm_du}) with $u^i$, we find another Lagrange multiplier:
\begin{align}
2\Lambda_0={}& {\frac 1{c^2}}\left[u^i{\frac {\delta{\cal L}^{\rm m}}
{\delta u^i}} - \nu{\frac {\delta{\cal L}^{\rm m}}{\delta\nu}} - \rho - p\right.\nonumber\\
{}&\left.+ J\nu\left(\frac{1}{2}\omega^2 - \left(\frac{1}{c^2}u^i\dot{N}_i\right)^2\right) + u^i
{\frac {\partial{\cal V}}{\partial u^i}}\right].\label{Lambda0}
\end{align}
Here again we cannot put equal zero the first two terms on the right-hand side
of (\ref{Lambda0}); they also should be inserted from the variation of (\ref{Lem1}).

Finally, let us notice that the material Lagrangian ``on-shell'' (i.e., after
making use of the field equations) reads:
\begin{align}\label{onLm}
{\cal L}^{\rm m} = p + \nu{\frac {\delta{\cal L}^{\rm m}}{\delta\nu}} - {\cal V}.
\end{align}

\section{Canonical Noether current tensor for matter}

The general form of the canonical energy-momentum tensor for the matter part is
\begin{align}\label{canEMTm}
{\stackrel{\rm m}\Sigma}{}_i{}^j := {\frac {\partial{\cal L}^{\rm m}}{\partial
\partial_j\Psi^A}}\,\partial_i\Psi^A - \delta_i{}^j\,{\cal L}^{\rm m}.
\end{align}
Here $\Psi^A$ are all dynamical variables. The matter Lagrangian (\ref{Lm})
depends on the derivatives of $N^i$ as well as on $\Lambda_1$, $s$, and $X$. The derivatives
with respect to the velocity of the director field are computed from 
(\ref{dLk_ddN}) and (\ref{dLp_ddN}):
\begin{align}
{\frac {\partial{\cal L}^{\rm m}}{\partial\partial_k N^i}} =
{\frac {\partial{\cal L}^{\rm k}}{\partial\partial_k N^i}} +
{\frac {\partial{\cal L}^{\rm p}}{\partial\partial_k N^i}} =
- J\nu\,u^k\,P^j{}_i\,\dot{N}_j -\,{\frac {\partial{\cal V}}
{\partial\partial_k N^i}}.\label{dLm_ddN}
\end{align}
On the other hand, from the constraint Lagrangian we find
\begin{align}
{\frac {\partial{\cal L}^{\Lambda}}{\partial\partial_i \Lambda_1}} = 
-\,\nu\,u^i,\qquad {\frac {\partial{\cal L}^{\Lambda}}{\partial\partial_i s}}
= \Lambda_2\,u^i,\qquad {\frac {\partial{\cal L}^{\Lambda}}{\partial
\partial_i X}} =  \Lambda_3\,u^i.\label{dLlsx}
\end{align}
Consequently, we have
\begin{align}
{\frac {\partial{\cal L}^{\rm m}}{\partial\partial_i N^k}}\,\partial_j N^k 
&= -\,u^i\,J\nu\,P^l{}_k\,\dot{N}_l\partial_j N^k -\,{\frac {\partial{\cal V}}
{\partial\partial_i N^k}}\partial_j N^k,\label{dLddN}
\end{align}
and
\begin{align}
{\frac {\partial{\cal L}^{\rm m}}{\partial\partial_i \Lambda_1}}\partial_j
\Lambda_1+{\frac {\partial{\cal L}^{\rm m}}{\partial\partial_i s}}\partial_j 
s + {\frac {\partial{\cal L}^{\rm m}}{\partial\partial_i X}}\partial_j X 
&= u^i\left(-\nu\partial_j\Lambda_1\right.\nonumber\\
&\left.+ \Lambda_2\partial_j s + \Lambda_3
\partial_j X\right).\label{dLdlsx}
\end{align}
In order to compute the right-hand side of (\ref{dLdlsx}), we can use 
(\ref{dLl_du}). The latter equation yields, with the help of (\ref{Lambda0})
and (\ref{Lambda5}):
\begin{align}
-\nu\partial_j\Lambda_1 + \Lambda_2\partial_j s + \Lambda_3\partial_j X 
= {\frac {u_j}{c^2}}\left[\nu{\frac {\delta{\cal L}^{\rm m}}{\delta\nu}}
+ p + \rho + {\frac J2}\nu\omega^2\right]\nonumber\\
+ P_j{}^k\left[J\nu\left(P_l{}^n\partial_k N^l - {\frac 1{c^2}}\,u^n\dot{N}_k
\right)\dot{N}_n + {\frac {\delta{\cal L}^{\rm m}}{\delta u^k}} + {\frac 
{\partial{\cal V}}{\partial u^k}}\right]\nonumber\\ 
- N_j{\frac {u^k}{c^2}}\left[{\frac {\delta{\cal L}^{\rm m}}
{\delta N^k}} - \partial_n\left(J\nu u^n P_k{}^l\dot{N}_l\right) 
+ {\frac {\delta{\cal V}}{\delta N^k}}\right]\label{dlsx}.
\end{align}
Substituting (\ref{onLm}), (\ref{dLddN}), (\ref{dLdlsx}), and (\ref{dlsx})
into (\ref{canEMTm}), we find explicitly the canonical energy-momentum tensor of 
the liquid crystal:
\begin{align}\label{canEMTfin}
{\stackrel{\rm m}\Sigma}{}_i{}^j= {\stackrel{\rm F}T}{}_i{}^j + 
u^j\,{\cal P}_i-P_i{}^j\,p^{\rm eff}.
\end{align}
Here the Frank deformation stress tensor is
\begin{align}\label{Femt}
{\stackrel{\rm F}T}{}_i{}^j := -\,{\frac {\partial{\cal V}}{\partial
\partial_j N^k}}\partial_i N^k + \delta_i^j\,{\cal V},
\end{align}
the effective pressure $p^{\rm eff}$ reads
\begin{align}
p^{\rm eff}:=p +\nu{\frac {\delta{\cal L}^{\rm m}}{\delta\nu}},\label{Peff} 
\end{align}
and the relativistic 4-momentum density of the fluid is
\begin{align}
{\cal P}_i:=&\, {\frac 1{c^2}}\,u_i\left(\rho -\frac{1}{2}J\nu\omega^2\right)\nonumber\\
+& P_i{}^k\left[-\,{\frac {J\nu}{c^2}}\dot{N}_k\,u^l\dot{N}_l + {\frac {\delta
{\cal L}^{\rm m}}{\delta u^k}} + {\frac {\partial{\cal V}}
{\partial u^k}}\right]\nonumber\\
-& N_i{\frac {u^k}{c^2}}\left[{\frac {\delta{\cal L}^{\rm m}}{\delta N^k}} 
- \partial_n\left(J\nu u^n P_k{}^l\dot{N}_l\right) + {\frac {\delta{\cal V}}
{\delta N^k}}\right].\label{Pmom}
\end{align}
As usual, we have to compute the variational derivatives of the matter
Lagrangian with respect to $\nu, u^i, N^i$ from the electromagnetic
part of the total Lagrangian (\ref{Lem1}). 

\subsection{Canonical spin of liquid crystal}

Denoting all the fields in the system by the symbol $\Psi^A$, that carries a
``multi-index'' $A$, with $(\rho_{ij})^A{}_B$ as the generators of the Lorentz 
algebra for these fields, the spin of the system is defined by
\begin{align}
{\stackrel {\rm m}S}_{ij}{}^k = {\frac {\partial{\cal L}^{\rm m}}{\partial(\partial_k
\Psi^A)}}\,(\rho_{ij})^A_B\Psi^B.\label{spin0}
\end{align}
Since the matter Lagrangian does not depend on the derivatives of $u^i$,
the spin of the system is straightforwardly computed with the help of 
(\ref{dLm_ddN}):
\begin{align}
{\stackrel {\rm m}S}_{ij}{}^k = N_{[i}\,{\frac {\partial{\cal L}^{\rm m}}{\partial\partial_k N^{j]}}} = 
- J\nu\,u^k N_{[i}P^l{}_{j]}\,\dot{N}_l - N_{[i}\,{\frac {\partial{\cal V}}
{\partial\partial_k N^{j]}}}.\label{spin}
\end{align}
Using the results of the Appendix \ref{explicitfrank}, we can write down the
variational derivatives for the Frank potential (\ref{VF}). They read:
\begin{align}
{\frac {\partial{\cal V}}{\partial u^i}} &= \,-\,{\frac {K_2}{c^2}}
\,(\partial_iN_j - \partial_jN_i)(\dot{N}^j - u^k\partial^jN_k)\nonumber\\
&- \frac{1}{{c^2}}(K_2 - K_3)(N^p\partial_pN_i)(N^q\partial_qN^k)u_k,\label{dVdu}\\
\frac{\partial {\cal V}}{\partial\partial_jN^i}&=K_1(\partial_kN^k)\delta^j_i + K_2\,P^j_kP^l_i
(\partial^kN_l - \partial_lN^k)\nonumber\\
&+ (K_2 - K_3)\,N^jP_i^k(N^p\partial_pN_k),\\
{\frac {\delta{\cal V}}{\delta N^i}}&=\partial_j\left[K_1(\partial_kN^k)\delta^j_i + K_2\,P^j_kP^l_i
(\partial^kN_l - \partial_lN^k)\right.\nonumber\\
&\left.\hspace{1cm}+ (K_2 - K_3)\,N^jP_i^k(N^p\partial_pN_k)\right]\nonumber\\ 
{}&+(K_2 - K_3)(\partial_iN^p)
(N^k\partial_kN_q)P^q_p.\label{eqVF}
\end{align}
As a consequence, the Frank stress tensor (\ref{Femt}) is 
\begin{align}
{\stackrel{\rm F}T}{}_i{}^j={}&-K_1(\partial_k N^k)(\partial_i N^j)+{\frac 12}K_1\delta_i^j\left(\partial_{k}N^k\right)^2\nonumber\\
&- K_2(\partial_iN_k)P^j_pP^k_q(\partial^pN^q-\partial^qN^p)\nonumber\\
&-(K_2 - K_3)(\partial_iN_k)N^jP_l^k(N^p\partial_pN^l)\nonumber\\
&+ {\frac 12}K_2\delta_i^j\left(\epsilon^{klm}N_{k}\partial_{l}N_{m}\right)^2\nonumber\\
&- {\frac 12}K_3\delta_i^j\left(\epsilon_{pmk}N^{m}\epsilon^{kln}
\partial_{l}N_{n}\right)^2,\label{TFtot}
\end{align}
and the spin density tensor of matter (\ref{spin}) reads
\begin{align}
{\stackrel {\rm m}S}_{ij}{}^k =& -\,J\nu\,u^k N_{[i}P^l{}_{j]}\,\dot{N}_l 
- K_1N_{[i}\,\delta_{j]}^k\,\partial_l N^l\nonumber\\
& -\,K_2P^{kn}N_{[i}P^l{}_{j]}(\partial_n N_l - \partial_l N_n)\nonumber\\
&- (K_2 - K_3)N^k N_{[i}P^l{}_{j]} N^n\partial_n N_l.\label{spin3}
\end{align}

\subsection{Balance equation of the angular momentum of matter}

Let us check that the angular momentum balance equation is satisfied for the
open material system under consideration. This balance equation follows from
the Noether theorem and it reads \cite{obukhov1}:
\begin{align}\label{angular0}
{\stackrel{\rm m}\Sigma}{}_{[ij]} + \partial_k\,{\stackrel {\rm m}S}_{ij}{}^k
= -\,{\frac {\delta{\cal L}^{\rm m}}{\delta\Psi^A}}\,(\rho_{ij})^A_B\Psi^B.
\end{align}
In our system, we have two vector fields, $u^i$ and $N^i$, and the 
corresponding Lorentz generators are $(\rho^i{}_j)^p_q = 
\delta^{[i}_{q}\delta^{p}_{j]}$. As a result, the right-hand side of
(\ref{angular0}) explicitly reads
\begin{align}\label{rhs}
-\,{\frac {\delta{\cal L}^{\rm m}}{\delta u^p}}\,(\rho_{ij})^p_q
\,u^q - {\frac {\delta{\cal L}^{\rm m}}{\delta N^p}}\,(\rho_{ij}
)^p_q\,N^q = {\frac {\delta{\cal L}^{\rm m}}{\delta u^{[i}}}
\,u_{j]} + {\frac {\delta{\cal L}^{\rm m}}{\delta N^{[i}}}\,N_{j]}.
\end{align}

{}From (\ref{canEMTfin}) we derive the first term on the left-hand side
of the balance equation (\ref{angular0}):
\begin{align}\label{canasym}
{\stackrel{\rm m}\Sigma}{}_{[ij]} = {\stackrel{\rm F}T}{}_{[ij]} 
+ {\cal P}_{[i}\,u_{j]}.
\end{align}
Denote
\begin{align}
\Phi_k:= {\frac {\delta{\cal L}^{\rm m}}{\delta N^k}} - \partial_n\left(J\nu u^n 
P_k{}^l\dot{N}_l\right) + {\frac {\delta{\cal V}}{\delta N^k}},\label{B}
\end{align}
then the field equation for the director (\ref{EqN}) is recast into
\begin{equation}
\pi_i{}^k\,\Phi_k = 0.\label{EqN2}
\end{equation}
Let us consider the last term in (\ref{canasym}). By making use of (\ref{Pmom}),
we have
\begin{align}
{\cal P}_{[i}\,u_{j]} &= {\frac {\delta{\cal L}^{\rm m}}{\delta u^{[i}}}\,u_{j]} 
+ {\frac {\partial{\cal V}}{\partial u^{[i}}}\,u_{j]}\nonumber\\
&- J\nu\dot{N}_{[i}{\frac{u_{j]}u^k}{c^2}}\,\dot{N}_k - N_{[i}{\frac {u_{j]}u^k}{c^2}}\,\Phi_k \nonumber\\
&= {\frac {\delta{\cal L}^{\rm m}}{\delta u^{[i}}}\,u_{j]} + {\frac {\partial
{\cal V}}{\partial u^{[i}}}\,u_{j]} + J\nu\,\dot{N}_{[i}P^k{}_{j]}\dot{N}_k\nonumber\\ 
&- N_{[i}\Phi_{j]} + N_{[i}\pi_{j]}{}^k\Phi_k.\label{can2}
\end{align}
We used the definition of the projector (\ref{pi}) that yields
\begin{align}
-\,{\frac {u_{j}u^k}{c^2}} = \pi_j{}^k - \delta_j^k - N_j N^k
\end{align}
to transform the two last terms on the first line in (\ref{can2}).

In view of the field equation (\ref{EqN2}) the last term in (\ref{can2})
vanishes, and we find
\begin{align}
{\cal P}_{[i}\,u_{j]} =& {\frac {\delta{\cal L}^{\rm m}}{\delta u^{[i}}}
\,u_{j]} + {\frac {\partial{\cal V}}{\partial u^{[i}}}\,u_{j]} 
+ J\nu\,\dot{N}_{[i}P^k{}_{j]}\dot{N}_k - N_{[i}\Phi_{j]}\nonumber\\
=& {\frac {\delta{\cal L}^{\rm m}}{\delta u^{[i}}}\,u_{j]} + {\frac {\delta
{\cal L}^{\rm m}}{\delta N^{[i}}}\,N_{j]} + {\frac {\partial{\cal V}}{\partial 
u^{[i}}}\,u_{j]}\nonumber\\
&+ {\frac {\delta{\cal V}}{\delta N^{[i}}}\,N_{j]}+\,J\nu\,\dot{N}_{[i}P^k{}_{j]}\dot{N}_k\nonumber\\
&+ N_{[i}\partial_n\left(J\nu u^n P_{j]}{}^k\dot{N}_k\right).\label{can3}
\end{align}
Recalling that $\dot{N}_i = u^n\partial_n N_i$, we see that
the last line reduces to the total divergence
\begin{align}\label{can4}
J\nu{}&\dot{N}_{[i}P^k{}_{j]}\dot{N}_k\nonumber\\
{}&+ N_{[i}\partial_n\left(J\nu u^n P_{j]}{}^k
\dot{N}_k\right) = \partial_n\left(J\nu u^n N_{[i}P_{j]}{}^k\dot{N}_k\right).
\end{align}

After these preparations, we can find the left-hand side of the balance
equation (\ref{angular0}):
\begin{align}
{\stackrel{\rm m}\Sigma}{}_{[ij]} + \partial_k\,{\stackrel {\rm m}S}_{ij}{}^k 
&={\frac {\delta{\cal L}^{\rm m}}{\delta u^{[i}}}\,u_{j]} + 
{\frac {\delta{\cal L}^{\rm m}}{\delta N^{[i}}}\,N_{j]}
+\,{\stackrel{\rm F}T}{}_{[ij]} + {\frac {\partial{\cal V}}{\partial 
u^{[i}}}\,u_{j]}\nonumber\\
&+ {\frac {\delta{\cal V}}{\delta N^{[i}}}\,N_{j]} 
- \partial_k\left(N_{[i}\,{\frac {\partial{\cal V}}
{\partial\partial_k N^{j]}}}\right).\label{angular2}
\end{align}
The last four terms in (\ref{angular2}) disappear in view of the Lorentz invariance of the Frank potential
${\cal V}$ that results, via the Noether theorem, in the following identity (see appendix \ref{explicitfrank}):
\begin{align}
{\stackrel{\rm F}T}{}_{[ij]} + {\frac {\partial{\cal V}}{\partial 
u^{[i}}}\,u_{j]} + {\frac {\delta{\cal V}}{\delta N^{[i}}}\,N_{j]} 
- \partial_k\left(N_{[i}\,{\frac {\partial{\cal V}}
{\partial\partial_k N^{j]}}}\right) = 0.\label{angular3}
\end{align}
Finally, using the identity (\ref{angular3}) in (\ref{angular2}), 
the angular momentum balance equation for matter can be recast into
\begin{align}
{\stackrel{\rm m}\Sigma}{}_{[ij]} + \partial_k\,{\stackrel {\rm m}S}_{ij}{}^k 
= {\frac {\delta{\cal L}^{\rm m}}{\delta u^{[i}}}\,u_{j]} + 
{\frac {\delta{\cal L}^{\rm m}}{\delta N^{[i}}}\,N_{j]}.\label{angular33}
\end{align}
Accordingly, by comparing (\ref{angular33}) with (\ref{rhs}) we indeed verify
the identity (\ref{angular0}) of the angular momentum for the open 
material system.

\section{Canonical energy-momentum tensor for the electromagnetic field inside the medium}

The canonical energy-momentum tensor for the electromagnetic field is obtained with the usual definition \cite{obukhov1} applied to the electromagnetic Lagrangian ${\cal L}^{\rm em}$ in (\ref{emlagrangian}). If we consider the 4-potential covector $A_i$ as the fundamental electrodynamical variable, then the electromagnetic canonical tensor ${\stackrel {\rm em}\Sigma}_{i}{}^{j}$ reads 
\begin{align}
{\stackrel{\rm em}\Sigma}{}_i{}^j:={}&{\frac {\partial{\cal L}^{\rm em}}{\partial
\partial_iA_k}}\,\partial_jA_k - \delta_j^i\,{\cal L}^{\rm em}\\
                                ={}& {\stackrel{\rm M}\Sigma}{}_i{}^j+H^{kj}(\partial_{k}A_{i}),\label{canonicalmink}
\end{align}
where ${\stackrel{\rm M}\Sigma}{}_i{}^j$ is the usual Minkowski tensor for the electromagnetic field in matter, given by
\begin{align}
{\stackrel{\rm M}\Sigma}{}_i{}^j:=-F_{ik}H^{jk}+\frac{1}{4}\delta_i{}^jF_{kl}H^{kl}.\label{defminkkk}
\end{align}
Now, using the expression (\ref{Hab}) in (\ref{canonicalmink}) and (\ref{defminkkk}), we get explicitly
\begin{align}
{\stackrel{\rm M}\Sigma}{}_i{}^j={}&\frac{1}{\mu_0}\left(\mu^{-1}_{\perp} + \Delta\mu^{-1}\right)
\left[-\,F^{jk}F_{ik} + {\frac 14}\delta_i^j F^{kl}F_{kl}\right]\nonumber\\
&+\,{\frac {1}{\mu_0c^2}}\left(\varepsilon_{\perp} -\mu^{-1}_{\perp} - \Delta\mu^{-1}\right)\left[
-\,F^{jk}u_k F_{il}u^l\right.\nonumber\\
&\left.\hspace{2.6cm}+ {\frac 12}\delta_i^j (F_{kl}u^l)^2 + u^j F_{ik}F^{kl}u_l
\right]\nonumber\\
& +\frac{1}{\mu_0}\Delta\mu^{-1}\left[-\,F^{jk}N_k F_{il}N^l + {\frac 12}
\delta_i^j (F_{kl}N^l)^2\right.\nonumber\\
&\left.\hspace{2cm}+ N^j F_{ik}F^{kl}N_l\right]\nonumber\\
& +\,{\frac 1{\mu_0c^2}}\left(\Delta\varepsilon + \Delta\mu^{-1}\right)(F_{pq}N^p u^q)
\left[-\,{\frac 12}\delta_i^j (F_{kl}N^k u^l)\right.\nonumber\\
&\hspace{2cm}\left.- u^j F_{in}N^n + N^j F_{in}u^n\right],\label{TM}
\end{align}
which is the explicit expression for the Minkowski tensor of the field inside the liquid crystal.

\subsection{Balance equation for the angular momentum of the electromagnetic field}

From the general definition \cite{obukhov1}, the spin density ${\stackrel {\rm em}S}_{ij}{}^{k}$ of the electromagnetic part of the system is given by,
\begin{align}
{\stackrel {\rm em}S}_{ij}{}^{k}={}\frac{\partial{\cal L}^{\rm em}}{\partial(\partial_{k}A_{m})}(\rho_{ij})_{m}{}^{l}A_{l} 
             ={}H^{k}{}_{[i}A_{j]}\neq0.\label{electrospin2}
\end{align}
Now we are in position to evaluate the angular momentum balance equation for the electromagnetic part of the system, which has the same form as the one for the matter part (\ref{angular0}). Taking the antisymmetric part of (\ref{canonicalmink}) and using the expression (\ref{electrospin2}), together with the Maxwell equations without sources, $\partial_jH^{ij}=0$, we see that the left-hand side of the identity for electromagnetic angular momentum is simply given by the antisymmetric part of the Minkowski tensor: 
\begin{align}
{\stackrel{\rm em}\Sigma}{}_{[ij]}+\partial_k{\stackrel {\rm em} S}_{ij}{}^k
=& {\stackrel{\rm M}\Sigma}{}_{[ij]},
\end{align}
where
\begin{align}
{\stackrel{\rm M}\Sigma}{}_{[ij]}={}&u_{[j} F_{i]k}\left[{\frac 
1{\mu_0c^2}}\left(\varepsilon_{\perp} - \mu^{-1}_{\perp} - \Delta\mu^{-1}\right)F^{kl}u_l\right.\nonumber\\
{}&\left.\hspace{1.2cm}- {\frac 1{\mu_0c^2}}\left(\Delta\varepsilon + \Delta\mu^{-1}\right)(F_{pq}
N^p u^q)\,N^k\right]\nonumber\\
&\,+\,N_{[j} F_{i]k}\left[\frac{1}{\mu_0}\Delta\mu^{-1} F^{kl}N_l\right.\nonumber\\
{}&\left.\hspace{0.6cm}+ {\frac 1{\mu_0c^2}}\left(\Delta
\varepsilon + \Delta\mu^{-1}\right)(F_{pq}N^p u^q)\,u^k\right].\label{angularbalanceelectro}
\end{align}
We see that the electromagnetic canonical energy-momentum tensor as well as the Minkowski tensor are not symmetric. However, it is not surprising that the right-hand side of (\ref{angularbalanceelectro}) is not equal to zero since the electromagnetic Lagrangian inside matter ${\cal L}^{\rm em}$ describes an open system.

On the other hand, computing the variations of ${\cal L}^{\rm em}$ in (\ref{Lem1}) with respect to the material variables yields, 
\begin{align}\label{dLedu}
{\frac {\delta{\cal L}^{\rm em}}{\delta u^i}} &= {\frac 1{\mu_0c^2}}\left(
\varepsilon_{\perp} - \mu^{-1}_{\perp} - \Delta\mu^{-1}\right)F_{ik}F^{kl}u_l\nonumber\\
&- {\frac 1{\mu_0c^2}}
\left(\Delta\varepsilon + \Delta\mu^{-1}\right)(F_{pq} N^p u^q)\,F_{ik}N^k,\\
{\frac {\delta{\cal L}^{\rm em}}{\delta N^i}} &= \frac{1}{\mu_0}\Delta\mu^{-1}
\,F_{ik}F^{kl}N_l\nonumber\\
&+{\frac 1{\mu_0c^2}}\left(\Delta\varepsilon +\Delta
\mu^{-1}\right)(F_{pq}N^p u^q)\,F_{ik}u^k.\label{dLedN}
\end{align}
Comparing (\ref{dLedu}) and (\ref{dLedN}) with (\ref{angularbalanceelectro}), we immediately
verify the correct balance equation for the electromagnetic angular momentum part of the system,
\begin{align}
{\stackrel{\rm em}\Sigma}{}_{[ij]}+\partial_k{\stackrel {\rm em} S}_{ij}{}^k= {\frac {\delta{\cal L}^{\rm em}}
{\delta u^{[i}}}\,u_{j]} + {\frac {\delta{\cal L}^{\rm em}}{\delta N^{[i}}}
\,N_{j]}.\label{angularM}
\end{align}
This is in perfect agreement with the general Noether identity (\ref{angular0}).

\section{Total canonical energy-momentum tensor}

The complete system of material medium plus electromagnetic field is described by the total Lagrangian
\begin{align}
{\cal L}:={\cal L}^{\rm m} + {\cal L}^{\rm em}.\label{totallag}
\end{align}
As a result, the total canonical energy-momentum tensor of the closed system is given by
\begin{align}
\Sigma_i{}^j:= {\stackrel{\rm m}\Sigma}{}_i{}^j + 
{\stackrel{\rm em}\Sigma}{}_i{}^j,\label{TT}
\end{align}
with the electromagnetic part ${\stackrel {\rm em} \Sigma}{}_i{}^j$ given in (\ref{canonicalmink}) and (\ref{TM}) and the material part ${\stackrel {\rm m} \Sigma}{}_i{}^j$ given in (\ref{canEMTfin}).

In order to find an explicit expression for the total canonical energy-momentum tensor $\Sigma_i{}^j$, we first need to evaluate the variations ${\delta{\cal L}^{\rm m}}/{\delta \nu}, {\delta{\cal L}^{\rm m}}/{\delta u^i}$ and ${\delta{\cal L}^{\rm m}}/{\delta N^i}$ and then insert them in (\ref{Peff}) and (\ref{Pmom}). For this aim, we take into account the equations of motion of the material variables
\begin{align}
{\frac {\delta{\cal L}^{\rm m}}{\delta \nu}} + {\frac {\delta{\cal L}^{\rm em}}
{\delta \nu}} &= 0,\label{Eqnu}\\
{\frac {\delta{\cal L}^{\rm m}}{\delta u^i}} + {\frac {\delta{\cal L}^{\rm em}}
{\delta u^i}} &= 0,\label{Equ}\\
{\frac {\delta{\cal L}^{\rm m}}{\delta N^i}} + {\frac {\delta{\cal L}^{\rm em}}
{\delta N^i}} &= 0,\label{EqNN}
\end{align}
from where we clearly see that the variations of the matter Lagrangian are exactly the negative of the variations of the electromagnetic Lagrangian, which we have already explicitly computed in (\ref{dLedu}) and (\ref{dLedN}). In addition, the variation of the 
electromagnetic Lagrangian with respect to the particle number density $\nu$, explicitly yields
\begin{align}
{\frac {\delta{\cal L}^{\rm em}}{\delta \nu}} =&\,-\,{\frac 12}\left(
\varepsilon_0\,{\frac {\partial\varepsilon}{\partial\nu}}\,{\cal E}^2 + {\frac 
1{\mu_0\mu_{\perp}^2}}\,{\frac {\partial\mu_{\perp}}{\partial\nu}}\,{\cal B}^2\right)\nonumber\\
&\,+  {\frac 12}\left(\varepsilon_0\,{\frac {\partial\Delta\varepsilon}
{\partial\nu}}\,({\cal E}_i N^i)^2 - {\frac 1{\mu_0}}\,{\frac {\partial
\Delta\mu^{-1}}{\partial\nu}}\,({\cal B}_i N^i)^2\right),\label{dLednu}
\end{align}
where we defined the 4-vectors electric field ${\cal E}_{i}$ and magnetic field ${\cal B}^{i}$ in (\ref{4e}) and (\ref{4b}), respectively. 

The variation (\ref{dLednu}) enters in the expression of the ``effective'' pressure (\ref{Peff}), which include the terms describing the electrostriction and magnetostriction effects. Then, replacing the negative of the three variations (\ref{dLedu}), (\ref{dLedN}), and (\ref{dLednu}) into (\ref{canEMTfin}), (\ref{Peff}), and (\ref{Pmom}), we explicitly obtain
\begin{align}
{\stackrel{\rm m}\Sigma}{}_i{}^j =& \,{\stackrel{\rm F}T}{}_i{}^j+ \,u^j{\widehat{\cal P}}{}_i -P_i{}^j\,p^{\rm eff} +\frac{1}{c^2}N_iu^ju^k\phi_k \nonumber\\
& \ -\frac{1}{\mu_0 c^2}\left(\varepsilon_{\perp} - \mu^{-1}_{\perp} - \Delta\mu^{-1}\right)u^jP_i{}^kF_{kl}F^{lm}u_m \nonumber\\
& \ +\frac{1}{\mu_0 c^2}\Delta\mu^{-1} N_iu^ju^kF_{kl}F^{lm}N_m \nonumber\\
& \ +\frac{1}{\mu_0 c^2}(\Delta\varepsilon+\Delta\mu^{-1}) u^jP_i{}^kF_{kl}N^l(F_{pq}N^pu^q),\label{explicitmatter}
\end{align}
where we denoted 
\begin{align}
\widehat{\cal P}_i &:= {\frac 1{c^2}}\,u_i\left(\rho - J\nu\omega^2/2\right) 
+ P_i{}^k\left[-\,{\frac {J\nu}{c^2}}\dot{N}_k\,u^l\dot{N}_l  
+ {\frac {\partial{\cal V}}{\partial u^k}}\right],\label{Phat}\\
\phi_i &:= \partial_k\left(J\nu u^k P_i{}^j\dot{N}_j\right) 
- {\frac {\delta{\cal V}}{\delta N^i}}.\label{phi}
\end{align}
The definitions of ${\stackrel F T}{}_i{}^j$ and $p^{\rm eff}$ are given in (\ref{Femt}) and (\ref{Peff}), respectively. Finally, if we consider (\ref{canonicalmink}), (\ref{TM}), and (\ref{explicitmatter}), we can write the total canonical tensor (\ref{TT}) explicitly, which reads
\begin{align}
\Sigma_i{}^j =&\,u^j{\widehat{\cal P}}{}_i + (-\,\delta^j_i
+ u^ju_i/c^2)\,p^{\rm eff} + {\stackrel{\rm F}T}{}_i{}^j +\frac{1}{c^2}N_iu^ju_k\phi^k,\nonumber\\
&+ \frac{1}{\mu_0}\left(\mu^{-1}_{\perp} + \Delta\mu^{-1}\right)
\left[-\,F^{jk}F_{ik} + {\frac 14}\delta_i^j F^{kl}F_{kl}\right]\nonumber\\
& +\,{\frac {1}{\mu_0 c^2}}\left(\varepsilon_{\perp} - \mu^{-1}_{\perp} - \Delta\mu^{-1}\right)\left[
-\,F^{jk}u_k F_{il}u^l\right.\nonumber\\
&\left.\hspace{2.7cm}+\left({\frac 12}\delta_i^j -{\frac {1}{c^2}}u_i u^j\right) (F_{kl}u^l)^2\right]\nonumber\\
&+\frac{1}{\mu_0}\Delta\mu^{-1}\left[-\,F^{jk}N_k F_{il}N^l + {\frac 12}
\delta_i^j(F_{kl}N^l)^2\right.\nonumber\\
&\left.\hspace{1.7cm}+\,N^j F_{ik}F^{kl}N_l + \frac{1}{c^2}N_iu^ju^k F_{kl}F^{lm}N_m\right]\nonumber\\
& +\,{\frac 1{\mu_0 c^2}}\left(\Delta\varepsilon + \Delta\mu^{-1}\right)
(F_{pq}N^p u^q)\left[N^j F_{in}u^n\right.\nonumber\\
&\left.\hspace{2cm}+\left(-\,{\frac 12}\delta_i^j + {\frac {1}{c^2}}
u_i u^j\right)(F_{kl}N^k u^l) \right]\nonumber\\
&+H^{kj}(\partial_{k}A_{i}).\label{Ttotall}
\end{align}

Since we already checked that the angular momentum balance equations are fulfilled both for the matter and electromagnetic parts of the total closed system in (\ref{angular33}) and (\ref{angularM}), respectively, it is obvious that if we add both equations, then the angular balance equation for the total system will be also valid,
\begin{align}\label{angularT}
\Sigma_{[ij]} + \partial_k\,S_{ij}{}^k = 0,
\end{align}
where
\begin{align}
S_{ij}{}^k:={\stackrel {\rm m}S}_{ij}{}^k+{\stackrel {\rm em}S}_{ij}{}^k.\label{totalspin}
\end{align}
It is worthwhile to notice that the right-hand side of (\ref{angularT})
vanishes since the total system is closed; however, the total
energy-momentum tensor (\ref{Ttotall}) is not symmetric, since the spin 
density of the system (\ref{totalspin}) is nontrivial.

\section{Fully explicit energy-momentum conservation law}

The total system under consideration, composed of a relativistic liquid crystal plus electromagnetic field, is a \textit{closed system}. There are no external fields present like $J^i_{\rm ext}$ and therefore the total canonical energy-momentum tensor (\ref{Ttotall}) of the system is conserved: 
\begin{align}
\partial_j\Sigma_i{}^j={}&0.
\end{align}

Notice that in the expressions (\ref{spin}), (\ref{electrospin2}), and (\ref{Ttotall})--(\ref{totalspin}), we have obtained the total energy-momentum tensor of the system and the identity of total angular momentum of the system, but without using a specific expression for the potential function ${\cal V}={\cal V}(N^i,\partial_jN^i,u^i)$. As a consequence, all the latter expressions are valid for an anisotropic uniaxial diamagnetic and dielectric medium, with any internal dynamics for the 4-director field $N^i$. By considering the expression (\ref{VF}) for the Frank potential ${\cal V}$, one can derive explicit expressions of the derivatives ${\delta{\cal V}}/{\delta u^i}$, ${\partial{\cal V}}/{\partial(\partial_j N^i)}$, ${\delta{\cal V}}/{\delta N^i}$ and of the tensors ${\stackrel {\rm F}T}_i{}^j$, ${\stackrel {\rm m}S}_{ij}{}^k$. This is done in detail in Appendix \ref{explicitfrank}.

In order to obtain an explicit expression of this conservation law, we can insert (\ref{Phat}), (\ref{phi}), (\ref{dVdu}), (\ref{eqVF}), and (\ref{TFtot}) into (\ref{Ttotall}). Due to the macroscopic Maxwell equations (\ref{macromax}), the gauge non-invariant term in (\ref{Ttotall}) vanishes and we finally obtain
\begin{align}
\partial_j{\stackrel {\rm c}\Sigma}_i{}^j+\partial_j{\stackrel {\rm f}\Sigma}_i{}^j=0.\label{conservlaw}
\end{align}
Here the non-symmetric energy-momentum tensor ${\stackrel {\rm c}\Sigma}_i{}^j$ only depends on the material variables, as if the relativistic liquid crystal were in isolation,
\begin{align}
{\stackrel {\rm c}\Sigma}_i{}^j:=&{\frac 1{c^2}}\,u_iu^j\left(\rho
-\frac{1}{2}J\nu\ \omega^k\omega_k\right)-P_i{}^jp^{\rm eff}\nonumber\\
&-{\frac {1}{c^2}}P_i{}^ku^j\left[K_2\,(\partial_kN_l - \partial_lN_k)(\dot{N}^l - u^p\partial^lN_p)\right.\nonumber\\
&\left.\hspace{0.6cm}+J\nu\dot{N}_k\,u^l\dot{N}_l +(K_2 - K_3)(N^p\partial_pN_k)(N^q\partial_qN^l)u_l\right]\nonumber\\ 
&-K_1(\partial_k N^k)(\partial_i N^j) - K_2(\partial_iN_k)P^j_pP^k_q(\partial^pN^q-\partial^qN^p)\nonumber\\
&-\,(K_2 - K_3)(\partial_iN_k)N^jP_l^k(N^p\partial_pN^l)\nonumber\\
&+{\frac 12}K_1\delta_i{}^j\left(\partial_{k}N^k\right)^2 + {\frac 12}
K_2\delta_i{}^j\left(\epsilon^{klm}N_{k}\partial_{l}N_{m}\right)^2\nonumber\\
{}&- {\frac 12}K_3\delta_i{}^j\left(\epsilon_{pmk}N^{m}\epsilon^{kln}
\partial_{l}N_{n}\right)^2\nonumber\\
&+\frac{1}{c^2}N_iu^j\left[J\nu u_ku^l\partial_l\left(P_j{}^k\dot{N}^j\right)+K_1u^k\partial_k(\partial_mN^m)\right.\nonumber\\
{}&\left.\hspace{0.6cm}+ (K_3 - K_2)u^k(\partial_kN^p)(N^l\partial_lN_q)P_p{}^q\right],\label{Sc}
\end{align}
except for some electromagnetic terms inside the effective pressure $p^{\rm eff}$, which describe possible electrostriction and magnetostriction effects:
\begin{align}
p^{\rm eff}={}&p+\,{\frac 12}\nu\left(
\varepsilon_0\,{\frac {\partial\varepsilon}{\partial\nu}}\,{\cal E}^2 + {\frac 
1{\mu_0\mu_{\perp}^2}}\,{\frac {\partial\mu_{\perp}}{\partial\nu}}\,{\cal B}^2\right)\nonumber\\
{}&-{\frac 12}\nu\left(\varepsilon_0\,{\frac {\partial\Delta\varepsilon}
{\partial\nu}}\,({\cal E}_i N^i)^2 - {\frac 1{\mu_0}}\,{\frac {\partial
\Delta\mu^{-1}}{\partial\nu}}\,({\cal B}_i N^i)^2\right).
\end{align}
The second term in (\ref{conservlaw}) corresponds to the 4-divergence of a non-symmetric ``field'' energy-momentum tensor ${\stackrel {\rm f}\Sigma}_i{}^j$, which contains electromagnetic field terms coupled to the material variables:
\begin{align}
{\stackrel {\rm f}\Sigma}_i{}^j:={}&\frac{1}{\mu_0}\left(\mu^{-1}_{\perp} + \Delta\mu^{-1}\right)
\left[- F_{ik}F^{jk} + {\frac 14}\delta_i^j F^{kl}F_{kl}\right]\nonumber\\
& +\,{\frac {1}{\mu_0 c^2}}\left(\varepsilon_{\perp} - \mu^{-1}_{\perp} - \Delta\mu^{-1}\right)\left[
-\,F^{jk}u_k F_{il}u^l\right.\nonumber\\
&\hspace{2cm}+\left.\left({\frac 12}\delta_i^j -{\frac {1}{c^2}}u_i u^j\right) (F_{kl}u^l)^2\right]\nonumber\\
& +\,\frac{1}{\mu_0}\Delta\mu^{-1}\left[-\,F^{jk}N_k F_{il}N^l + {\frac 12}
\delta_i^j(F_{kl}N^l)^2\right.\nonumber\\
&\left.\hspace{1.1cm}+N^j F_{ik}F^{kl}N_l + \frac{1}{c^2}N_iu^ju^k F_{kl}F^{lm}N_m\right]\nonumber\\
& +\,{\frac 1{\mu_0 c^2}}\left(\Delta\varepsilon + \Delta\mu^{-1}\right)
(F_{pq}N^p u^q)\left[N^j F_{in}u^n\right.\nonumber\\
&\left.\hspace{0.8cm}+\left(-\,{\frac 12}\delta_i^j + {\frac {1}{c^2}}
u_i u^j\right)(F_{kl}N^k u^l)\right].\label{Sf}
\end{align}

\section{Relativistic director dynamics}

Let us analyze the equations of motion for the director $N^i$. Rewriting equations (\ref{EqN}) or (\ref{EqN2}), we have
\begin{align}
\pi_i{}^j\left[\partial_k\left(J\nu\,u^k\,P^l{}_j\,\dot{N}_l\right)+h_j\right] = 0,\label{EqN3}
\end{align}
where we defined
\begin{align}
h_j:={}&-{\frac{\delta{\cal V}}{\delta N^j}}+\frac{\partial {\cal L}^{\rm em}}{\partial N^j},\label{totalmolecular}
\end{align}
as the total {\it 4-molecular field}, since its spatial components reduce, in the non-relativistic limit, to the standard ``molecular field'' \cite{lisin1,degennes}. In Sec. \ref{norellimit} we study this limit in more detail. In order to better interpret the dynamics of $N^i$, we can contract (\ref{EqN3}) with $\epsilon^{pqi}N_q$, make use of (\ref{epsomega}) and the continuity equation (\ref{C1}), to obtain
\begin{align}\label{dotomega}
J\nu\,\pi^i{}_j\,\dot{\omega}^j=-\epsilon^{ijk}N_jh_k.
\end{align}

From (\ref{dotomega}) we see that the 4-molecular field $h_i$ is responsible for the ``torques'' and changes in the 4-director $N^i$ of the liquid crystal. We can identify two contributions to the 4-molecular field $h_i:={\stackrel {\rm F}h}_i+{\stackrel {\rm em}h}_i$. One is the Frank deformation 4-molecular field
\begin{align}
{\stackrel {\rm F}h}_i:={}&-\frac{\delta {\cal V}}{\delta N^i}\\
                      ={}&\partial_j\left[K_1(\partial_kN^k)\delta_i{}^j + K_2\,P_k{}^jP_i{}^l
(\partial^kN_l - \partial_lN^k)\right.\nonumber\\
&\left.\hspace{1cm}-(K_3 -K_2)\,N^jP_i{}^k(N^p\partial_pN_k)\right]\nonumber\\ 
{}&+(K_3 - K_2)(\partial_iN^p)(N^k\partial_kN_q)P_p{}^q,
\end{align}
which describes the changes in $N^i$ caused by the deformations of the liquid crystal itself, and the other contribution is the electromagnetic 4-molecular field
\begin{align}
{\stackrel {\rm em}h}_i:={}&\frac{\partial {\cal L}^{\rm em}}{\partial N^i}\\
                      ={}&\frac{1}{\mu_0}\Delta\mu^{-1}
\,F_{ik}F^{kl}N_l\nonumber\\
{}&+{\frac 1{\mu_0c^2}}\left(\Delta\varepsilon +\Delta
\mu^{-1}\right)(F_{pq}N^p u^q)\,F_{ik}u^k,\label{herel}
\end{align}
which describes the influence of the dynamical electromagnetic field on the orientation of the 4-director.

With the knowledge of  (\ref{conservlaw}), together with Eqs. (\ref{dotomega}) for $N^i$, the continuity equation $\partial_i(\nu u^i)=0$ in (\ref{C1}) for $u^i$, $\nu$ and the Maxwell equations (\ref{macromax}) for $F_{ij}$, we can completely determine the dynamics and evolution of this system, composed of the relativistic liquid crystal with anisotropic optical properties interacting with the electromagnetic field. 

\section{Non-relativistic limit}\label{norellimit}

Let us study the dynamics of the liquid crystal in the non-relativistic limit, 
when the motion of the fluid is such that $|\bm{v}|\ll c$. In particular, this 
approximation can be applied when the liquid crystal is at rest in the 
laboratory frame. In the non-relativistic limit, we expect consistency with 
the earlier results \cite{lisin1,lisin2,degennes,stewart}, but first we need
some technical preparations. The 4-velocity reads
\begin{align}
u^i=  \gamma(1, \bm{v}),\label{4velnonrel}
\end{align}
and therefore we have for the components of the projector:
\begin{align}
P_a{}^b={}&\delta_a{}^b + {\frac {v_av^b}{c^2}},\qquad
P_a{}^0={} \gamma^2{\frac {v_a}{c^2}},\\
P_0{}^a={}&-\gamma^2v^a,\qquad\hspace{0.5cm}
P_0{}^0={} \gamma^2{\frac {v^2}{c^2}}.
\end{align}
Hereafter, the 3-dimensional indices are raised and lowered by the Euclidean metric; in particular, $\bm{v} = v^a$, $v_a = \delta_{ab}v^b$, $v^2 = v^av_a = \delta_{ab}v^av^b$, etc.
For the skew-symmetric tensor (\ref{eps}) we find explicitly
\begin{align}
\epsilon^{abc}={}& \gamma{\stackrel{\circ}\epsilon}{}^{abc},\qquad \hspace{0.4cm}\epsilon^{0ab}= \gamma{\stackrel{\circ}\epsilon}{}^{abc}{\frac {v_c}{c^2}},\\
\epsilon_{abc}={}& -\gamma{\stackrel{\circ}\epsilon}{}_{abc},\qquad \epsilon_{0ab}= \gamma{\stackrel{\circ}\epsilon}{}_{abc}v^c.
\end{align}
Taking into account the orthogonality conditions (\ref{defni})--(\ref{connn1}), the 4-director $N^i$ reads in components
\begin{align}
N^i=\left({\frac {(\bm{v}\cdot\bm{n})}{c^2}},\bm{n}\right).\label{4dirnonrel}
\end{align}
Notice that the 3-vector $\bm{n}$, with Cartesian components $n^a$, recovers its normalization $\bm{n}^2 = \delta_{ab}n^an^b=1$ only in the non-relativistic limit. In general,
it satisfies $\bm{n}^2 = 1 + (\bm{v}\cdot\bm{n})^2/c^2$.

\subsection{Non-relativistic director dynamics to zeroth order in $v/c$}

First we note that $N_j\dot{\omega}^j=0$ in (\ref{dotomega}) and therefore we obtain
\begin{align}
J\nu P_j{}^i\dot{\omega}^j=-\epsilon^{ijk}N_jh_k.\label{dotomega2}
\end{align}
Then, inserting (\ref{4velnonrel})--(\ref{4dirnonrel}) into (\ref{dotomega2}), we find the non-relativistic equation for $n^a$ to zeroth order in $v/c$ 
\begin{align}
J\nu\,\dot{\bm\omega}{}_{\circ} = \bm{n}\times\bm{h}{}_{\circ}.\label{dotomeganorel}
\end{align}
Here,
\begin{align}
\dot{\bm\omega}{}_{\circ} = \bm{n}\times\frac{\partial^2\bm{n}}{\partial t^2},
\end{align}
is the angular acceleration of a liquid crystal fluid element to zeroth order in $v/c$. The right-hand side of (\ref{dotomeganorel}) is determined by 
the molecular field, $\bm{h}{}_{\circ}:={\stackrel {\rm F}{\bm h}}{}_{\circ}+{\stackrel {\rm em}{\bm h}}{}_{\circ}$. The fluid part
\begin{align}
{\stackrel {\rm F}h}_a{}^{\circ}={}&(K_1- K_2)(\partial_a\partial_bn^b) + K_2\delta_{ab}\bm{\nabla}^2n^b
\nonumber\\ 
{}&+(K_3 - K_2)\delta_{ad}\ \partial_b(n^bn^c\partial_cn^d)\nonumber\\
{}&-(K_3 - K_2)\delta_{bd}(\partial_an^b)(n^c\partial_cn^d),\label{ha}
\end{align}
is the zeroth order Frank deformation molecular field (cf. the non-relativistic equation (2.147) of \cite{stewart}). The electromagnetic part of the molecular field can be computed by taking the non-relativistic limit to zeroth order in $v/c$ of (\ref{herel}), which explicitly reads
\begin{align}
{\stackrel {\rm em}{\bm h}}{}_{\circ}={}&\varepsilon_0\Delta\varepsilon(\bm{E}\cdot\bm{n})\bm{E}-\mu_0^{-1}\Delta\mu^{-1}(\bm{B}\cdot\bm{n})\bm{B}\nonumber\\
{}&+\mu_0^{-1}\Delta\mu^{-1} B^2\bm{n}.\label{henorel}
\end{align}
The result (\ref{henorel}) looks slightly different from the usual electromagnetic molecular field of the non-relativistic models
\cite{lisin2,degennes,stewart}, since we use the fields $\bm{E}$ and $\bm{B}$ as the independent fields and not $\bm{E}$ and $\bm{H}$. 
One can, however, easily recover the same formulas in a different disguise by making use of the inverse of the constitutive relations (\ref{c2}).
Notice that the last term in (\ref{henorel}) is proportional to the director $\bm{n}$ and therefore it does not contribute to the torque when replaced in the cross product of (\ref{dotomeganorel}). Therefore, we can ignore this last term and redefine the electromagnetic molecular field to zeroth order in $v/c$, as a sum of two terms, the electric molecular field ${\stackrel {\rm e}{\bm h}}{}_{\circ}$ and the magnetic molecular field ${\stackrel {\rm m}{\bm h}}{}_{\circ}$, given by
\begin{align}
{\stackrel {\rm e}{\bm h}}{}_{\circ}={}&\varepsilon_0\Delta\varepsilon(\bm{E}\cdot\bm{n})\bm{E},\\
{\stackrel {\rm m}{\bm h}}{}_{\circ}={}&-\frac{1}{\mu_0}\Delta\mu^{-1}(\bm{B}\cdot\bm{n})\bm{B}.
\end{align}

Collecting together all the results of this section, the dynamics of the director $\bm{n}$ in the nonrelativistic limit, to zeroth order in $v/c$, is described by
\begin{align}
J\nu\,\bm{n}\times\frac{\partial^2 \bm{n}}{\partial t^2}=\bm{n}\times\bm{h}^{\rm F}_{\circ}+\bm{n}\times\bm{h}^{\rm e}_{\circ}+\bm{n}\times\bm{h}^{\rm m}_{\circ},\label{nonrelvector}
\end{align}
where the molecular field has independent contributions from the Frank deformations and the interactions with electric and the magnetic fields.

\subsection{Non-relativistic solutions}

Let us assume that the electromagnetic field vanishes, so that
$\bm{h}^{\rm e}_{\circ}=\bm{h}^{\rm m}_{\circ}=\bm{0}$. 

Stewart in Sec. 2.5 of \cite{stewart} describes some exact solutions of
the equations of motion. One is an obvious constant director solution; the 
other is the static spherical solution and the twist 
solution [with the local coordinates $\bm{x}=(x^1, x^2, x^3)$]:
\begin{align}
n^a &= {\frac {x^a}{r}},\label{sp}\\
n^a &= \left(\cos\theta, \sin\theta, 0\right),
\qquad\theta = c_1x^3 + c_2.\label{ts}
\end{align}

Here we notice that the static twist solution (\ref{ts}) can be generalized to a dynamical ``plane wave" solution:
\begin{align}
n^a &= \left(\cos\Theta, \sin\Theta, 0\right),\qquad \Theta = c_0t + c_1x^3 + c_2,\label{ts2}
\end{align}
where $c_0,c_1,c_2$ are arbitrary constants. In this case, we explicitly have, 
\begin{align}
\frac{\partial^2 \bm{n}}{\partial t^2}={}&-(c_0)^2\bm{n},\\
\bm{h}^{\rm F}_{\circ}={}& -K_2(c_1)^2\bm{n},
\end{align}
and thus \eqref{nonrelvector} is satisfied. In the literature, the wave
solutions of the full nonlinear equations of motion has attracted
some attention (see \cite{lisin1,lisin2,kamen}).

\section{Summary and Discussion: Abraham-Minkowski controversy}

In this paper, we have constructed a complete relativistic 
Lagrangian theory of a nematic liquid crystal. Our results provide a consistent 
relativistic model for a medium with anisotropic optical properties in 
interaction with the electromagnetic field. In particular, in such a framework 
one can study the problem of the proper description of the energy and momentum 
of light in anisotropic media. This should shed light on the long standing 
Abraham-Minkowski controversy, traditionally discussed only for isotropic media.

We have generalized the earlier non-relativistic model \cite{lisin1,lisin2} 
and the variational model of an ideal relativistic 
fluid \cite{obukhov1} and derived a complete theory of the nematic liquid
crystal medium and its interaction with the electromagnetic field.
We have derived the nonlinear equations of motion for the liquid crystal
fluid, and explicitly verified the total energy, momentum, and angular momentum balance laws, which arise as consequences of the Noether
theorem from the invariance of the (field plus matter) system under 
spacetime translations and Lorentz transformations, respectively.
In this work the general formalism is presented in full detail. We will analyze the solutions
and applications separately. The analysis of the properties of electromagnetic 
waves in the moving liquid crystal will be also considered elsewhere. 

As we have seen, liquid crystals are an interesting example of continuous media
with microstructure. The ``internal'' degrees of freedom of such a medium is
represented by the director vector field $N^i$ assigned to every material
point of the fluid. This field gives rise to a nontrivial spin of the medium
(\ref{spin3}). As a consequence, the total energy-momentum tensor (\ref{TT}) of the 
closed system of the medium plus the electromagnetic field is not symmetric.
This asymmetry is crucial for the validity of the Noether identities. 

In this paper, we have neglected dissipation. The general formalism developed
here is applicable to any moving medium with uniaxial anisotropic properties. In particular, one should have in mind possible astrophysical applications 
\cite{balakin}, where our model provides an explicit dynamical mechanism for
the description of a physical medium with uniaxial anisotropy. As concerns
the liquid crystals which belong to the class of moving uniaxial media, they
are characterized by a nontrivial dissipation. In this sense, our theory has
limited applicability and it should be considered only as a first
step in constructing a full realistic physical model. The relativistic 
mechanics of dissipative fluids has a long and controversial history, with
the first attempts going back to Eckart \cite{Eckart} and Landau-Lifshitz
\cite{LLF}. Later an essential improvement was achieved in the works of
Israel and Stewart \cite{Israel1,Israel2}. However, these models used a 
phenomenological approach with numerous \textit{ad hoc} assumptions, and it seems more
appropriate to use the approach of Carter \cite{Carter,Priou,livrev}, which 
generalizes the variational principle by replacing the Lagrangian with the
so-called {\it master function} that systematically takes into account the
irreversible viscosity and thermal effects. The development of the Carter 
type approach for dissipative media with microstructure will constitute the
next step for the construction of the variational theory of liquid crystals.
An important feature is that the Noether type identities play a central
role in Carter's approach. 

Our current results contribute to the discussion of the energy and momentum problem of the electromagnetic field in a medium. 
In particular, our analysis clearly demonstrates that for the case of an uniaxial anisotropic medium it is the total 
energy-momentum tensor of the coupled system (matter plus field) that is 
important for the understanding of the balance of the momentum and angular 
momentum. For the case of an isotropic medium, this was pointed out more that 40 years ago by Penfield and Haus \cite{penfieldhaus1,penfieldhaus1967} and more recently in \cite{obukhov1,dielectricslab,pfeifer}.

Notwithstanding this general fact of the importance of the {\it total} energy-momentum, one can split it into a ``matter'' and a ``field'' part in many different ways. Specifically, by expressing (\ref{Ttotall}) in terms of $N^i$, $u^i$, $F_{ij}$, permittivities and impermeabilities, we have shown that there exists a split of the form (\ref{Sc}) plus (\ref{Sf}). The purely matter part (\ref{Sc}) does not depend explicitly on the field strength $F_{ij}$, except for the terms present in effective pressure. Therefore, it could be identified with what is sometimes called a ``kinetic'' energy-momentum \cite{barnett2010,barnettloudon2}. The corresponding field part (\ref{Sf}) has a form which is more complicated than the usual Abraham tensor for isotropic media \cite{obukhov1}, since the former involves not only the field $F_{ij}$, the 4-velocity $u^i$, the isotropic permittivity $\varepsilon$ and the permeability $\mu$, but also the 4-director $N^i$, its derivatives, and the anisotropies 
$\Delta\varepsilon$, $\Delta\mu^{-1}$. Despite the fact that its structure is different from that of Abraham, this tensor plays a role analogous to that of the traditional Abraham tensor and moreover reduces to it in the case $\Delta\varepsilon=\Delta\mu^{-1}=0$. 

However, the tensor (\ref{Sf}) is not symmetric. It was pointed out in 
\cite{obukhov1} that, for simple media where the 4-velocity $u^i$ is the only 
non-scalar field contained in the constitutive relation, the total 
energy-momentum tensor turns out to be the sum of a kinetic term plus the 
Abraham tensor. It was not clear whether the same is the case for anisotropic 
media. Our results now clarify this point. Another interesting problem would be 
to find a consistent definition of some kind of ``generalized Abraham 
tensor'', possessing the same properties of the Abraham tensor in the isotropic 
case, but valid in any type of medium. This would certainly improve our 
understanding of the Abraham-Minkowski controversy, which has been restricted 
only to simple media. Results in this direction will be analyzed in a 
forthcoming publication.

\appendix

\section{Explicit calculation of the Frank potential terms}\label{explicitfrank}

In this appendix the explicit derivatives of the Frank potential are computed and an 
explicit verification of the Noether identity (\ref{angular3}) is given. For the computation of the derivatives 
of the Frank potential (\ref{VF}), the following identities will be useful:
\begin{align}
\epsilon_{ijk}\epsilon^{pqk} \equiv{}& P^p{}_j P^q{}_i - P^p{}_i P^q{}_j,\label{epsPP}\\
P^i{}_j N^j \equiv{}& N^i,\\
N^l\epsilon^{pqr} \equiv{}& N^p\epsilon^{lqr} + N^q\epsilon^{plr} + N^r\epsilon^{pql},\label{Ne}
\end{align}
where $P^i{}_j$ and $\epsilon_{ijk}$ are defined in (\ref{projectoruu}) and (\ref{eps}), respectively. Then, using (\ref{epsPP})--(\ref{Ne}), we can prove these other useful relations:
\begin{align}
\epsilon_{ijk}N^j\epsilon^{kln}\partial_l N_n 
={}& P^n_i N^l\partial_l N_n,\label{K3red}\\
P^i{}_j\,\dot{N}^j={}& \epsilon^{ijn}\omega_j N_n,\label{dotNomega}\\
2N_{[k}\epsilon_{l]ij}\omega^i N^j ={}& -\,\epsilon_{kln}\omega^n=2N_{[k}P^j{}_{l]}\,\dot{N}_j. \label{epsomega}
\end{align}

We now are in condition to compute the contribution of the Frank potential to the dynamics of the fluid. For this, it is convenient to separate the potential \eqref{VF} into three pieces:
\begin{equation}
{\cal V}:={\cal V}_1+{\cal V}_2+{\cal V}_3.\label{frankparts}
\end{equation}

\subsection{The first elastic constant}

For the splay deformation elastic potential,
\begin{equation}
{\cal V}_1 = \frac{1}{2}K_1(\partial_i N^i)^2,\label{V111}
\end{equation}
we find
\begin{align}
{\frac {\partial{\cal V}_1}{\partial u^{i}}}={}&{\frac {\partial{\cal V}}{\partial N^{i}}}=0,\label{dV1}\\
{\frac {\partial{\cal V}_1}{\partial\partial_j 
N^{i}}}={}&K_1(\partial_k N^k)\,\delta^j_i,\\ 
{\frac {\delta{\cal V}_1}{\delta N^{i}}} ={}& {\frac {\partial{\cal V}_1}{\partial 
N^{i}}} - \partial_j\left({\frac {\partial{\cal V}_1}{\partial\partial_j N^{i}}}
\right) =- K_1\partial_i(\partial_k N^k).\label{eqV1}
\end{align}
Then, using (\ref{dV1})-(\ref{eqV1}) in (\ref{Femt}), we obtain
\begin{align}
{\stackrel{\rm F1}T}{}_i{}^j ={}& -\,K_1(\partial_k N^k)(\partial_i N^j) 
+ \delta_i^j{\cal V}_1,\label{temV1}\\
{\stackrel{\rm F1}T}{}_{[ij]}= {}&-\,K_1(\partial_k N^k)(\partial_{[i} N_{j]}).
\end{align}
Finally, let us calculate explicitly the terms:
\begin{align}
{\frac {\delta{\cal V}_1}{\delta N^{[i}}}\,N_{j]} =& -\,K_1N_{[j}
\partial_{i]}(\partial_k N^k),\label{expression111}\\
- \partial_k\left(N_{[i}\,{\frac {\partial{\cal V}_1}{\partial\partial_k N^{j]}}}\right)
=& -\,K_1(\partial_k N^k)(\partial_{[j} N_{i]})\nonumber\\
{}&-\,K_1N_{[i}\partial_{j]}(\partial_k N^k).\label{expression222}
\end{align}
Substituting all this into (\ref{angular3}), we verify the Noether identity for ${\cal V}_1$. 

\subsection{The second elastic constant}

The second term (twist deformation) in the Frank potential is given by
\begin{align}\label{V20}
{\cal V}_2 = {\frac 12}K_2\left(\epsilon^{ijk}N_{i}\partial_{j}N_{k}\right)^2.
\end{align}
Using the identities (\ref{epsPP}) and (\ref{Ne}), we can rewrite this as
\begin{align}
{\cal V}_2={}& {\frac 14}K_2\,\pi^i_k\pi^j_l(\partial_iN_j - \partial_jN_i)
(\partial^kN^l - \partial^lN^k) \\
={}& {\frac 12}K_2\left[P^i_kP^j_l(\partial_iN_j - \partial_jN_i)\partial^kN^l\right.\nonumber\\
{}&\left.\hspace{1cm}+ P^i_j(N^p\partial_pN_i)(N^q\partial_qN^j)\right].\label{V2}
\end{align}
The derivatives of the potential ${\cal V}_2$ with respect to the director
and its derivatives are straightforwardly calculated:
\begin{align}
{\frac {\partial{\cal V}_2}{\partial N^i}}={}&K_2(\partial_iN^p)
(N^k\partial_kN_q)P^q_p \nonumber\\
={}&K_2\,(\partial_iN^k)(N^p\partial_pN_k)\nonumber\\
{}&-{\frac {K_2}{c^2}}(\partial_iN^p)
u_p(N^q\partial_qN^k)u_k,\label{dV2dN}\\
{\frac {\partial{\cal V}_2}{\partial \partial_jN^i}}={}& K_2\left[P^j_kP^l_i(
\partial^kN_l - \partial_lN^k) + N^jP_i^k(N^p\partial_pN_k)\right]\nonumber\\
={}& K_2\left(\partial^jN_i -\partial_iN^j + N^jN^p\partial_pN_i\right)\nonumber\\
{}&-{\frac {K_2}{c^2}}\left[u^j(\dot{N}_i - u^k\partial_iN_k) - u_i(
\dot{N}^j - u^k\partial^jN_k)\right.\nonumber\\
{}&\left.\hspace{1.1cm}+ N^ju_i(N^p\partial_pN^k)u_k\right].\label{dV2ddN}
\end{align}
In addition, the derivative with respect to the velocity reads
\begin{align}
{\frac {\partial{\cal V}_2}{\partial u^i}} &= \,-\,{\frac {K_2}{c^2}}\left[
(\partial_iN_j - \partial_jN_i)(\dot{N}^j - u^k\partial^jN_k)\right.\nonumber\\
{}&\hspace{1.5cm}\left.+ (N^p\partial_pN_i)(N^q\partial_qN^k)u_k\right].\label{dV2du}
\end{align}
Substituting (\ref{dV2ddN}) into the definition (\ref{Femt}), we obtain the 
contribution of $K_2$ to the energy-momentum tensor 
\begin{align}
{\stackrel{\rm F2}T}{}_i{}^j &= \delta_i^j\,{\cal V}_2 - K_2(\partial_iN_k)
\left[P^j_pP^k_q(\partial^pN^q - \partial^qN^p)\right.\nonumber\\
{}&\left.\hspace{3.2cm}+N^jP_l^k(N^p\partial_pN^l)\right]\\
&= \delta_i^j\,{\cal V}_2 - K_2(\partial_iN_k)\left[\partial^jN^k - 
\partial^kN^j + N^j(N^p\partial_pN^k)\right]\nonumber\\ \label{Femt2}
&\quad + {\frac {K_2}{c^2}}(\partial_iN_k)\left[u^j(\dot{N}^k - u^l\partial^k
N_l)-u^k(\dot{N}^j - u^l\partial^jN_l)\right.\nonumber\\
&\hspace{2.5cm}+\left.N^j(N^p\partial_pN^l)u^ku_l\right].
\end{align}

In order to check the Noether identity (\ref{angular3}) for the ${\cal V}_2$ term, we first notice that
it can be identically recast into
\begin{align}\label{NoeF}
{\stackrel{\rm F2}T}{}_{[ij]} + {\frac {\partial{\cal V}_2}{\partial 
u^{[i}}}\,u_{j]} +  {\frac {\partial{\cal V}_2}{\partial N^{[i}}}\,N_{j]} 
- (\partial_kN_{[i})\,{\frac {\partial{\cal V}_2}{\partial\partial_k N^{j]}}}= 0.
\end{align}
Using (\ref{dV2ddN}), we derive an intermediate result:
\begin{align}
(\partial_kN_j)\,{\frac {\partial{\cal V}_2}{\partial \partial_kN^i}}={}&K_2
\left[(\partial^kN_j)(\partial_kN_i - \partial_iN_k)\right.\nonumber\\
{}&\left.\hspace{0.7cm}+ (N^p\partial_pN_j)(N^q\partial_qN_i)\right]\nonumber\\
{}&- {\frac {K_2}{c^2}}\left[\dot{N}_j(\dot{N}_i - u^k\partial_iN_k)\right.\nonumber\\ 
{}&\hspace{0.4cm}- u_i(\partial^kN_j)(\dot{N}_k - u^l\partial_kN_l)\nonumber\\
{}&\left.\hspace{0.25cm}+u_iu^k(N^p\partial_pN_j)(N^q\partial_qN_k)\right].\label{dNdV2dN}
\end{align}
It is straightforward to find the antisymmetric objects using (\ref{dV2dN}), 
(\ref{dV2du}), (\ref{Femt2}), and (\ref{dNdV2dN}):
\begin{align}
{\stackrel{\rm F2}T}{}_{[ij]}={}&-\,K_2(\partial^kN_{[i})\partial_{j]}N_k- {\frac {K_2}{c^2}}\dot{N}_{[j}(\partial_{i]}N_k)u^k\nonumber\\
{}&+K_2N_{[i}(\partial_{j]}N^k)(N^p\partial_pN_k)\nonumber\\
{}&-\,{\frac {K_2}{c^2}}\,u_{[i}(\partial_{j]} N_k)(\dot{N}^k - u^l
\partial^kN_l)\nonumber\\
{}&- {\frac {K_2}{c^2}}\,N_{[i}(\partial_{j]}N^k)u_k(N^q\partial_qN_l)u^l,\label{NoeV2a}\\
{\frac {\partial{\cal V}_2}{\partial u^{[i}}}\,u_{j]}={}& {\frac {K_2}{c^2}}
\,u_{[i}(\partial_{j]} N_k)(\dot{N}^k - u^l\partial^kN_l)\nonumber\\
{}&- {\frac {K_2}{c^2}}\,u_{[i}(\partial^k N_{j]})(\dot{N}_k - u^l\partial_kN_l)\nonumber\\ \label{NoeV2b}
{}&+\,{\frac {K_2}{c^2}}\,u_{[i}N^p(\partial_{|p|} N_{j]})(N^q\partial_qN_k)u^k,\\
{\frac {\partial{\cal V}_2}{\partial N^{[i}}}\,N_{j]}={}& -\,K_2N_{[i}
(\partial_{j]}N^k)(N^p\partial_pN_k)\nonumber\\
{}&+ {\frac {K_2}{c^2}}\,N_{[i}(\partial_{j]}N^k)
u_k(N^q\partial_qN_l)u^l,\label{NoeV2c}
\end{align}
and 
\begin{align}
(\partial_kN_{[i})\,{\frac {\partial{\cal V}_2}{\partial\partial_k N^{j]}}}={}&
\,K_2(\partial^kN_{[j})\partial_{i]}N_k-{\frac {K_2}{c^2}}\dot{N}_{[j}(\partial_{i]}N_k)u^k\nonumber\\
{}&-{\frac {K_2}{c^2}}\,u_{[i}(\partial^kN_{j]})(\dot{N}_k - u^l\partial_kN_l)\nonumber\\
{}&+{\frac {K_2}{c^2}}\,u_{[i}N^p(\partial_{|p|} N_{j]})(N^q\partial_qN_k)u^k.\label{NoeV2d}
\end{align}
Substituting all this into (\ref{NoeF}), we verify the Noether identity (\ref{angular3}) for the
second term ${\cal V}_2$. 

\subsection{The third elastic constant}

Analogously, we consider the third term (bend deformation) in the Frank potential, 
\begin{align}\label{V3}
{\cal V}_3 ={}&-\,{\frac 12}K_3\left(\epsilon_{ijk}N^{j}\epsilon^{kln}\partial_{l}
N_{n}\right)^2\\
={}& -\,{\frac 12}K_3\,P^j_i(N^p\partial_p N_j)(N^q\partial_q N^i),
\end{align}
where we have used (\ref{K3red}) to simplify the potential. It is worthwhile to
notice that this quadratic invariant has the same form as the last term in the $K_2$
potential (\ref{V2}). As a result, the corresponding derivatives of the 
potential (\ref{V3}) with respect to its arguments can be conveniently
extracted from the formulas (\ref{dV2dN}), (\ref{dV2ddN}), and (\ref{dV2du}). These 
derivatives are given by
\begin{align}
{\frac {\partial{\cal V}_3}{\partial N^i}}={}&-\,K_3(\partial_iN^p)
(N^k\partial_kN_q)P^q_p\\
={}& -\,K_3\,(\partial_iN^k)(N^p\partial_pN_k)\nonumber\\
{}&+{\frac {K_3}{c^2}}(\partial_iN^p)
u_p(N^q\partial_qN^k)u_k,\label{dV3dN}\\
{\frac {\partial{\cal V}_3}{\partial \partial_jN^i}}={}&-\,K_3\,N^jP_i^k(N^p
\partial_pN_k)\\ \label{dV3ddN}
={}&-\,K_3\,N^jN^p\partial_pN_i\nonumber\\
{}&+{\frac {K_3}{c^2}}\,N^ju_i(N^p\partial_pN^k)u_k,\\
{\frac {\partial{\cal V}_3}{\partial u^i}}={}& {\frac {K_3}{c^2}}
\,(N^p\partial_pN_i)(N^q\partial_qN^k)u_k.\label{dV3du}
\end{align}
The stress tensor (\ref{Femt}) for ${\cal V}_3$ reads 
\begin{align}
{\stackrel{\rm F3}T}{}_i{}^j={}&\delta_i^j\,{\cal V}_3 + K_3(\partial_iN_k)
N^jP_l^k(N^p\partial_pN^l)\\
={}&\delta_i^j\,{\cal V}_3 + K_3(\partial_iN_k) N^j(N^p\partial_pN^k)\nonumber\\
{}&-{\frac {K_3}{c^2}}(\partial_iN_k)N^j(N^p\partial_pN^l)u^ku_l.\label{Femt3}
\end{align}
In order to check the Noether identity for this last part of the potential ${\cal V}$, 
we observe that Eqs. (\ref{dV3dN})--(\ref{Femt3}) yield
\begin{align}\label{NoeV3a}
{\stackrel{\rm F3}T}{}_{[ij]}={}&-\,K_3N_{[i}(\partial_{j]}N^k)(N^p\partial_pN_k)\nonumber\\ 
{}&+ {\frac {K_3}{c^2}}\,N_{[i}(\partial_{j]}N^k)u_k(N^q\partial_qN_l)u^l,\\
{\frac {\partial{\cal V}_3}{\partial u^{[i}}}\,u_{j]}={}& -\,{\frac {K_3}{c^2}}
\,u_{[i}N^p(\partial_{|p|} N_{j]})(N^q\partial_qN_k)u^k,\label{NoeV3b}\\
{\frac {\partial{\cal V}_3}{\partial N^{[i}}}\,N_{j]}={}& K_3N_{[i}
(\partial_{j]}N^k)(N^p\partial_pN_k)\nonumber\\
{}&- {\frac {K_3}{c^2}}\,N_{[i}(\partial_{j]}N^k)
u_k(N^q\partial_qN_l)u^l,\label{NoeV3c}\\ \label{NoeV3d}
(\partial_kN_{[i})\,{\frac {\partial{\cal V}_3}{\partial\partial_k N^{j]}}}={}&-
{\frac {K_3}{c^2}}\,u_{[i}N^p(\partial_{|p|} N_{j]})(N^q\partial_qN_k)u^k.
\end{align}
Substituting all this into (\ref{NoeF}), we verify the Noether identity
(\ref{angular3}) for the third term ${\cal V}_3$.

\section*{Acknowledgments}

For YNO this work was partly supported by the German-Israeli Foundation for
Scientific Research and Development (GIF), Research Grant No.\ 1078-107.14/2009. 
G.R. acknowledges financial support for part of this research from FONDECYT, through project 1060939. 
T.R. acknowledges financial support from BECAS CHILE scholarship program.

\end{document}